\documentclass{emulateapj}
\usepackage{amssymb}
\usepackage{apjfonts}
\usepackage{graphicx}
\usepackage{natbib}
\usepackage{multirow}
\newcommand{\lya}{Ly~$\alpha$}

\def\CIV{C\,{\sc iv}}

\newcommand{\CIVab}{C{\sevenrm~IV}\,$\lambda\lambda$1548,1551}
\newcommand{\CIVwave}{C{\sevenrm~IV}\,$\lambda$1549}

\def\MgII{Mg\,{\sc ii}}
\def\MgIIwave{Mg\,{\sc ii}\,$\lambda$2798}

\newcommand{\MgIIab}{Mg{\sevenrm~II}\,$\lambda\lambda$2796,2803}

\newcommand{\OIIIb}{[O{\sevenrm\, III}]\,$\lambda$5007}

\def\vabs{$v_{\rm abs}$}
\def\zabs{$z_{\rm abs}$}
\def\zem{$z_{\rm em}$}
\def\kms{$\rm km\,s^{-1}$}
\def\ergs{${\rm erg\,s^{-1}}$}

 \font\sevenrm=cmr7 scaled 1000

\begin{document}
\title{Collective properties of quasar narrow associated absorption lines}
\shorttitle{Collective properties of quasar associated absorptions}
\shortauthors{Chen et al.}

\author{Zhi-Fu Chen\altaffilmark{1}, Da-Sheng Pan\altaffilmark{2}}

\altaffiltext{1}{School of Materials Science and Engineering, Baise University, Baise 533000, China; zhichenfu@126.com}

\altaffiltext{2}{Department of Information Technology, Guangxi Financial Vocational College, Nanning 530007, China}

\begin{abstract}
This paper statistically investigates the properties of \CIV\ and \MgII\ narrow absorption lines (NALs) to look for velocity cuts that can well constrain quasar-associated NALs. The coverage fraction ($f_c$) is defined as the ratio between the number of quasars exhibiting at least one detected absorber and the total number of quasars that can be used to detect absorptions with given criteria. We find that, for both \CIV\ and \MgII\ absorbers, both the number density of absorbers in given velocity intervals ($dn/d\beta$) and the $f_c$ show very significant excess at the low-velocity offset from the quasars, relative to the random occurrence that is expected for cosmologically intervening absorbers. These relative excess extensions for \MgII\ absorptions are not only evidently related to absorption strength but also to quasar luminosity, while they are mainly constrained within 2000 \kms\ no matter what quasar luminosity and absorption strength are. In addition, we find that the redshift number density ($dn/dz$) evolution of \MgII\ absorbers with \vabs\ $<2000$ \kms\ evidently differs from that with \vabs\ $>2000$ \kms. Turning to \CIV\ absorptions, the relative excess extensions of both $dn/d\beta$ and $f_c$ are mainly limited within \vabs$<4000$ \kms, and depend neither on absorption strength nor on quasar luminosity. And also, the absorbers with \vabs$<4000$ \kms\ show obviously different redshift number density evolution from those with \vabs$>4000$ \kms. We suggest velocity cuts of 4000 \kms\ and 2000 \kms\ to define quasar \CIV\ and \MgII\ associated NALs, respectively.
\end{abstract}

\section{Introduction}
A widely accepted event is that supermassive black holes (SMBHs) reside at the central region of all massive galaxies, mostly in the form of active galactic nuclei (AGNs) \cite[e.g.][]{2013ARA&A..51..511K}, and the mass of the SMBH ($M_{\rm BH}$) can reach up to $10^{10}~M_{\odot}$ \cite[e.g.,][]{2015Natur.518..512W}. The evolution behavior of star formation rate density within galaxies is similar to that of the black hole accretion rate density \cite[e.g.,][]{2010MNRAS.401.2531A,2014ARA&A..52..415M}. The black hole mass is tightly related with the stellar mass \cite[e.g.,][]{2013ARA&A..51..511K,2017ApJ...837L...8B} or stellar velocity dispersion \cite[e.g.,][]{2009MNRAS.397.1705G,2013ARA&A..51..511K,2015ApJ...809...20B,2017ApJ...838L..10B} of the host galaxy bulge. These signatures reveal a connection between the AGN activity and global properties of host galaxies, which is widely researched and debated in the last decades \cite[e.g.,][]{2000ApJ...539L...9F,2005Natur.433..604D,2008ApJ...681..931B,2012MNRAS.427.3103B,2013ARA&A..51..511K,2017MNRAS.468.3395M,2017A&A...601A..63W,2017MNRAS.469..295B}. However, the underlying physics is still controversial. For example, why, when and how the SMBHs and their host galaxies affect and/or regulate one another, which hampers us from completely comprehending the cosmic evolutions of SMBHs and galaxies.

SMBHs' accretion of surrounding gas is a foundational process in their lifetimes. The disk gas of AGNs accelerated by radiation pressure \cite[e.g.,][]{1995ApJ...454L.105M,2001MNRAS.326..916C,2007ApJ...661..693P}, thermal pressure \cite[e.g.,][]{2001ApJ...561..684K,2012MNRAS.422.1880O}, magnetocentrifugal forces \cite[e.g.,][]{2005ApJ...631..689E,2015ApJ...805...17F,2017MNRAS.465.1741C}, cosmic ray pressure \cite[e.g.,][]{1991A&A...245...79B,2013ApJ...777L..16B}, and/or a combination of them, then leaves off the central region in form of outflow/wind. The gas accretion and outflow are two important hands to regulate the global properties of SMBH and host galaxy. AGN outflows are proven to be an efficient process to halt the gas infall infinitely into both the galaxies and central SMBHs. These outflows are also a good transport tool that carries away the energy, matter and momentum, which could be injected into broad emission line region, narrow emission line region, host galaxy, circumgalactic medium (CGM), and intergalactic medium (IGM). These injections, fashionably called feedback, could quench star formation rate by removing gas off its location or heating gas up to very high temperature \cite[e.g.,][]{2012A&A...537L...8C,2012ApJ...745L..34Z,2015ApJ...799...82C,2016A&A...591A..28C,2017MNRAS.468.4956Z}, and on the other hand, are also usually invoked to explain the enhancement of star formation rate through compressing gas clouds \cite[e.g.,][]{2012MNRAS.427.2998I,2015ApJ...799...82C,2016MNRAS.455.4166B,2017MNRAS.468.4956Z}.

AGN outflows can be observed via blueshifted emission and absorption features against continuum emissions of central compact regions. Therein absorptions are more fashionable tool to probe global properties of outflows. Quasar absorptions often exhibit complex features, which can be roughly classified into broad absorption lines (BALs), mini-BALs and narrow absorption lines (NALs) based on line widths of their profiles, and can be also divided into associated and intervening absorptions in term of whether they are physically related to the quasar system or not. Complex features indicate that different type of absorption lines would reflect absorbers with different characteristics. BALs display smooth absorption troughs with velocity widths being larger than a few thousands \kms\ at depths $>10\%$ below the continuum, and are believed to be undoubtedly associated with quasars. NALs generally show sharp profiles with full width at half maximum (FWHM) being less than a few hundreds \kms\ and can be formed in a wide variety of medium, no matter what the relationship between the medium and quasar is. When compared to the detected BALs and NALs, the detected mini-BALs that have line widths between BALs and NALs are much rarer. Hence, the mini-BALs lack in-depth research and are poorly understood \cite[e.g.,][]{2007ApJ...660..152M,2009ApJ...696..924G,2010ApJ...724..762W,2011A&A...536A..49G,2013ApJ...775...14R,2016PASJ...68...48H,2016MNRAS.457.2665M,2017MNRAS.468.4539M}.

BALs and mini-BALs are usually blended and we find that it difficult to understand the properties of absorbers in greater detail. Resolved NALs, in contrast, are often observed in spectra even at a middle or low resolution. For example, unblended \CIVab\ and \MgIIab\ doublets \cite[e.g.,][]{2015ApJS..221...32C,2016MNRAS.462.2980C} are often imprinted on quasar spectra of the Sloan Digital Sky Survey \cite[][]{2000AJ....120.1579Y}, which have a resolution of $R=1300\sim2500$ \cite[][]{2015ApJS..219...12A}. Resolved NALs could lead us to realize the absorber's ionization level, gas density, metallicity, dynamical process, and so on. Therefore, NALs would be an important tool for investigating the physical conditions and environments of quasars.

Unlike BALs, it is not easy to diagnose which NALs are truly associated with quasars, since NALs with similar profiles can be formed in environments no matter whether or not they are related to quasars. The frequently used methods that infer quasar-associated NALs, mainly include: (1) line variability with time; (2) line profiles that are obviously smoother and broader than those mainly dominated by thermal motion; (3) partial coverage fraction of absorber to background emission source; (4) absorptions requiring strong radiation field or high gas density; (5) significantly statistical excess of absorbers compared to cosmologically intervening ones.

Quasar host galaxy, host galaxy CGM, outflow, and nearby galaxies within the same one cluster/group are able to produce associated absorptions, so one often expects a cluster distribution of NALs around quasars. However, due to distance galaxies that absorb the background, quasar photons randomly distribute in the foreground space of quasars, cosmologically intervening in the absorptions of quasars, and thus would not exhibit obvious excess relative to the random distribution. Thus, statistical analysis of absorber distributions is a practical tool to distinguish associated absorptions from intervening ones.

Large sky surveys are beneficial because they statistically limit the fraction of quasar-associated absorptions, especially the quasar spectroscopy of the Sloan Digital Sky Survey \cite[][]{2000AJ....120.1579Y} which has obtained more than 0.4 million unique quasar spectra \cite[e.g.,][]{2010AJ....139.2360S,2017A&A...597A..79P}. In this work, we will use the largest known quasar narrow-line absorption catalogs that contain both associated and intervening \MgIIab\ or \CIVab\ doublets to statistically look for empirical evidence that separates quasar-associated absorptions from intervening ones.

We describe the data sample in Section \ref{sect:datasample}, present the statistical properties of absorptions and discussions in Section \ref{sect:discussion}. The conclusions and summary are presented in Section \ref{sect:summary}. In this paper, we adopt the $\rm \Lambda CDM$ cosmology with $\rm \Omega_M=0.3$, $\rm \Omega_\Lambda=0.7$, and $\rm H_0=70~km~s^{-1}~Mpc^{-1}$.

\section{data sample}
\label{sect:datasample}
The Sloan Digital Sky Survey \cite[SDSS;][]{2000AJ....120.1579Y} is one of the most ambitious projects in astronomy, and uses a dedicated wide-field 2.5 telescope \cite[][]{2000AJ....120.1579Y}, located at Apache Point Observatory, New Mexico to map the universe. The SDSS obtained the first light in May 1998, and have collected more than 3 million spectra at a resolution of $R\approx2000$ from the regular survey operation in 2000. The Baryon Oscillation Spectroscopic Survey \cite[BOSS;][]{2013AJ....145...10D} is an important mapping of the third phase of the SDSS (SDSS-III), which utilized the original SDSS 2.5m telescope with updated multi-object fiber fed optical spectrographs \cite[][]{2013AJ....146...32S} to gather data in the main dark time from 2008 July to 2014 June and produce spectra in the range of 3600 \AA\ $< \lambda<$ 10400 \AA\ at a resolution of $R=1300\sim2500$ \cite[][]{2015ApJS..219...12A}.

Quasar spectra of the SDSS have been widely used to detect cold gas absorption features, such as \MgIIab, \CIVab\ doublets. These absorbing gases might be constrained by the quasars or a part of the quasars (associated absorptions), and might also be well beyond the gravitational bound (intervening absorptions). Most of the absorption line groups \cite[e.g.,][]{2005ApJ...628..637N,2011AJ....141..137Q,2013ApJ...763...37C,2013ApJ...779..161S,2013ApJ...770..130Z,2016MNRAS.463.2640R} mainly focused on intervening absorptions in their programs that systematically searched for metal absorptions on the SDSS quasar spectra. Using the first data release of the BOSS (the ninth data release of the SDSS), which includes $\rm87~822$ unique quasar spectra \cite[DR9Q;][]{2012A&A...548A..66P}, a series works of our absorption line group \cite[e.g.,][]{2014ApJS..210....7C,2014ApJS..215...12C,2015ApJS..221...32C,2016MNRAS.462.2980C} were designed to search for metal absorptions on quasar spectra, and have produced, up to today, the largest absorption catalogs that contain not only associated but also intervening \MgII\ or \CIV\ NALs. This provides us an excellent data set to statistically analyze the properties of quasar-associated NALs.

\cite{2015ApJS..221...32C,2016MNRAS.462.2980C} searched for \MgII\ or \CIV\ absorptions with strengths no smaller than 0.2 \AA\ at rest-frame and significance level larger than $2\sigma$ in the quasar spectra data redward of \lya\ emissions until red wings of \MgIIwave\ or \CIVwave\ emissions. In this work, we directly pick out data, including quasar and absorption system samples, from \cite{2015ApJS..221...32C,2016MNRAS.462.2980C} to complete our statistical analysis. Our selected criteria are as follows:
\begin{enumerate}
  \item The BAL (broad absorption line) flag \cite[][]{2012A&A...548A..66P} equals to 0, which indicates that the quasar spectra is not imprinted a BAL feature. The quasar with a ZWARNING $>0$ is also excluded, which flags bad fits in the redshift-fitting code \cite[][]{2012A&A...548A..66P}.
  \item Quasar redshifts that are determined by broad emission lines, especially by the asymmetric and/or possibly blueshifted \CIV\ emission lines, often exhibit a large uncertainty from 100 \kms\ to 3000 \kms\ with respect to those measured from narrow emission lines \cite[e.g.,][]{2010MNRAS.405.2302H,2011ApJS..194...45S}. \cite{2016ApJ...817...55S} recently uncovered that, after accounting for measurement errors, quasar redshifts measured by broad \MgII\ emission lines would show an uncertainty of $\sim$ 200 \kms\ when compared to the redshifts calculated by \OIIIb\ narrow emission lines. We limit quasars (here after \MgII\ quasar sample) with $0.4<z_{em}<1.1$ for the investigations of \MgII\ absorptions, so that both narrow \OIIIb\ and \MgII\ associated absorptions can be available by the BOSS spectra. Turning to the studies of \CIV\ absorptions, we only consider quasars (here after \CIV\ quasar sample) with $1.4<z_{em}<2.4$ so that both \MgII\ emissions and \CIV\ associated absorptions can be imprinted on the BOSS spectra. These limits of quasar sample could reduce the uncertainty introduced by quasar redshfits.

      The quasar redshift is critical for this study, which is one of the dominant reasons why we constrain our quasar samples with the above redshift ranges. The SDSS pipeline redshifts are highly accurate for most of the quasars, which are determined by a combination of multi eigenspectra \cite[][]{2012AJ....144..144B}. In order to correct for the bad pipeline redshifts, \cite{2017A&A...597A..79P} visually checked all objects included in the Quasar Data Release 12 (DR12Q) and modified the bad pipeline redshifts to \MgII\ based redshifts. In addition, for quasars with $z_{\rm em}<0.9$, we find that the differences between visual inspection redshifts of \cite{2017A&A...597A..79P} and \OIIIb\ based redshifts (in preparation) show a negligible mean offset of 22 \kms\ and a dispersion of 90 \kms, which suggest that the visual inspection redshifts provided by \cite{2017A&A...597A..79P} are robust for low redshift quasars. Therefore, we utilize the visual inspection redshifts provided by \cite{2017A&A...597A..79P} for the \MgII\ quasar sample. For the \CIV\ quasar sample, we directly adopt the \MgII\ based redshifts from \cite{2017A&A...597A..79P} when available, otherwise we utilize the visual inspection redshifts. The redshifts of the quasars included in these two subsamples are exhibited in Figure \ref{fig:zem}.
      \begin{figure}
      \centering
      \includegraphics[width=0.43\textwidth]{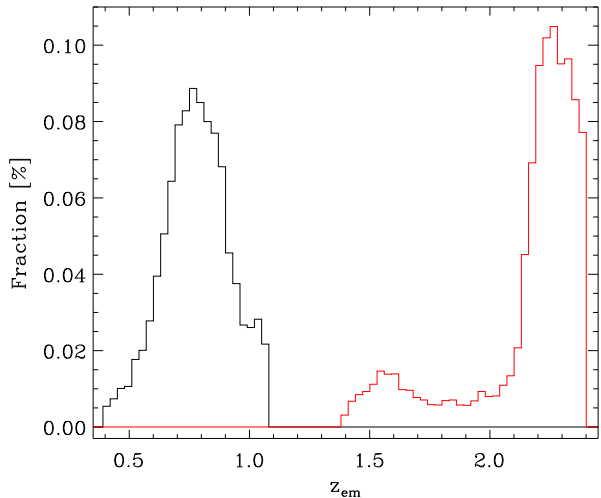}
      \caption{Black line is for the quasars used to investigate the properties of \MgII\ absorptions, and red line is for the quasars used to investigate the properties of \CIV\ absorptions. Y-axis represents the number of quasars that have been normalized by the total number of quasars.}
      \label{fig:zem}
      \end{figure}

  \item We cut off the absorptions with $W_r^{\lambda2796} \ge 0.3$ \AA\ and $W_r^{\lambda2796}/\sigma_{W_r^{\lambda2796}} \ge 3$ for \MgII\ doublets, and $W_r^{\lambda1548} \ge 0.3$ \AA\ and $W_r^{\lambda1548}/\sigma_{W_r^{\lambda1548}} \ge 3$ for \CIV\ doublets.
\end{enumerate}
In terms of the above limits, we take 9214 quasars and 1533 \MgII\ absorption doublets from \cite{2015ApJS..221...32C} for the studies of \MgII\ associated absorptions, and adopt 12156 quasars and 11648 \CIV\ absorption doublets from \cite{2016MNRAS.462.2980C} for the statistical analysis of \CIV\ associated absorptions.

\section{Statistical analysis and discussions}
\label{sect:discussion}
\subsection{Velocity offset distributions}\label{sect:Velocity_offset}
Assuming the difference between absorption and emission redshifts is originated from the relative motion of absorbers with respect to quasars, the velocity offset of absorbers relative to the quasar system can be derived via
\begin{equation}
\label{eq:vr}
\beta\equiv\frac{v_{abs}}{c} = \frac{(1+z_{em})^2 - (1+z_{abs})^2}{(1+z_{em})^2 + (1+z_{abs})^2},
\end{equation}
where $c$ is the speed of light, \zem\ is the quasar redshift, and \zabs\ is the absorber redshift. The absorptions of quasar local environments are expected to exhibit an evidently inequable $\beta$ distribution at $\beta\approx0$ when compared to the distribution of intervening absorptions ($\beta\gg0$). This is confirmed by many early works that obviously excessive numbers of absorbers with $\beta\approx0$ are over a random distribution \cite[e.g.,][]{1979ApJ...234...33W,1986ApJ...307..504F,2003ApJ...599..116V,2008MNRAS.386.2055N,2008MNRAS.388..227W,2016MNRAS.462.3285P}. The statistically significant excess of absorbers around quasars provides convenient cuts to distinguish associated absorptions from intervening ones. In the last two decades, we often use the conventional boundaries of 3000 \kms\ to assemble \MgII\ associated \cite[\vabs$<3000$ \kms; e.g.,][]{2006MNRAS.367..945Y,2008ApJ...679..239V,2012ApJ...748..131S,2014ApJ...794...66K} and intervening \cite[\vabs$>3000$ \kms; e.g.,][]{2005ApJ...628..637N} absorptions, and 5000 \kms\ to construct \CIV\ associated (\vabs$<5000$ \kms) and intervening (\vabs$>5000$ \kms) absorption samples.

The number density of absorbers ($dn/d\beta\equiv cdn/d\upsilon$) is defined as the number of absorbers per unit of velocity interval. The redshift range of the quasar sample and the wavelength coverage range of spectra may play a role to the distribution of the $dn/d\beta\equiv cdn/d\upsilon$, since some spectra of low-redshift quasars cannot be used to detect absorptions with high velocity offsets from quasars. The redshift distributions of the quasars included in our samples are shown in Figure \ref{fig:zem}. The BOSS quasar spectra wavelength range is from 3600 \AA\ to 10400 \AA. We find that there are $>99\%$ of the spectra of \MgII\ quasar sample can be used to detected \MgII\ absorption with $v_{\rm abs}>30000$ \kms, and $>95\%$ of the spectra of \CIV\ quasar sample can be used to detected \CIV\ absorption with $v_{\rm abs}>30000$ \kms.

\begin{figure*}
\centering
\includegraphics[width=0.47\textwidth]{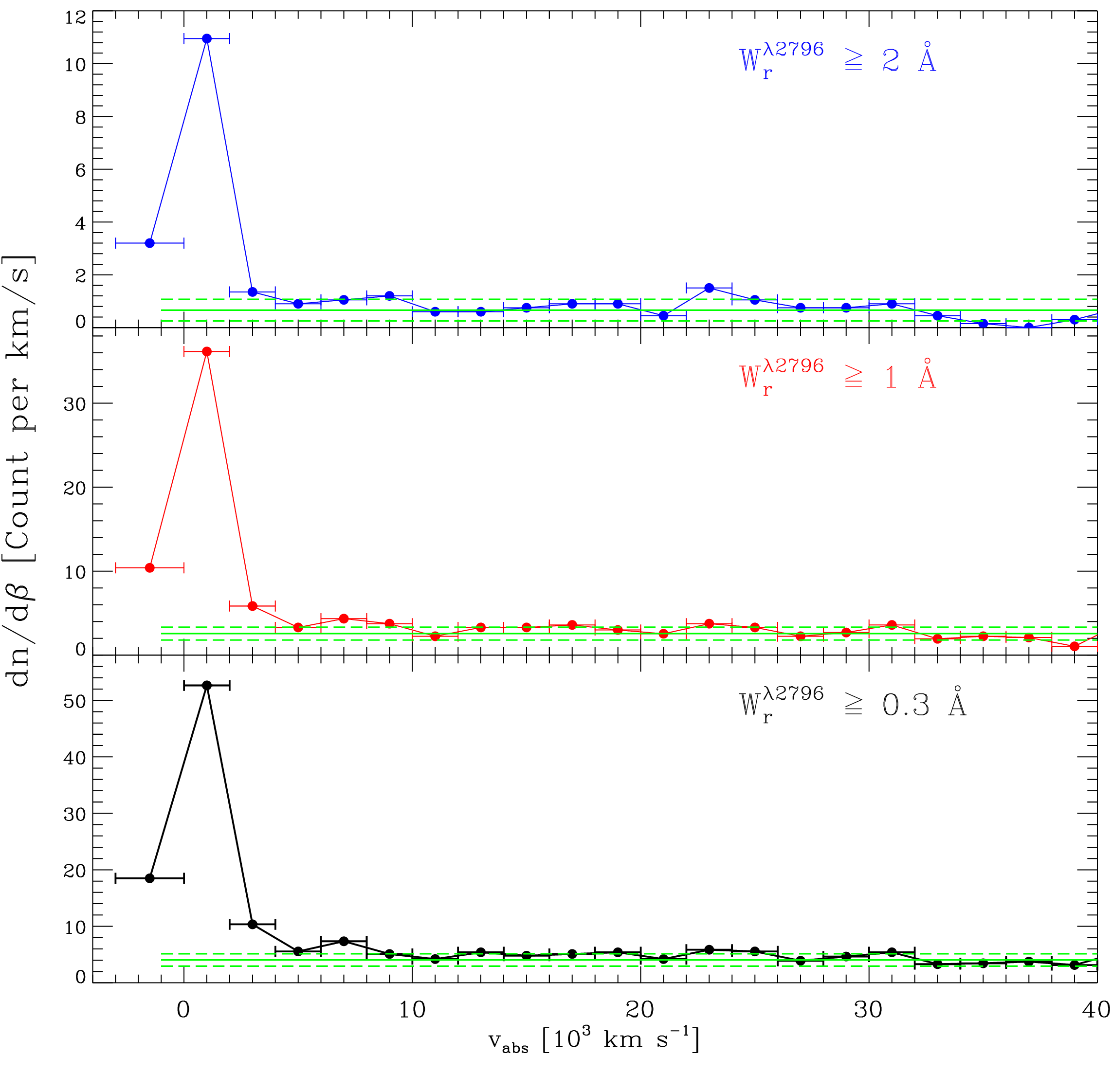}
\hspace{3ex}
\includegraphics[width=0.47\textwidth]{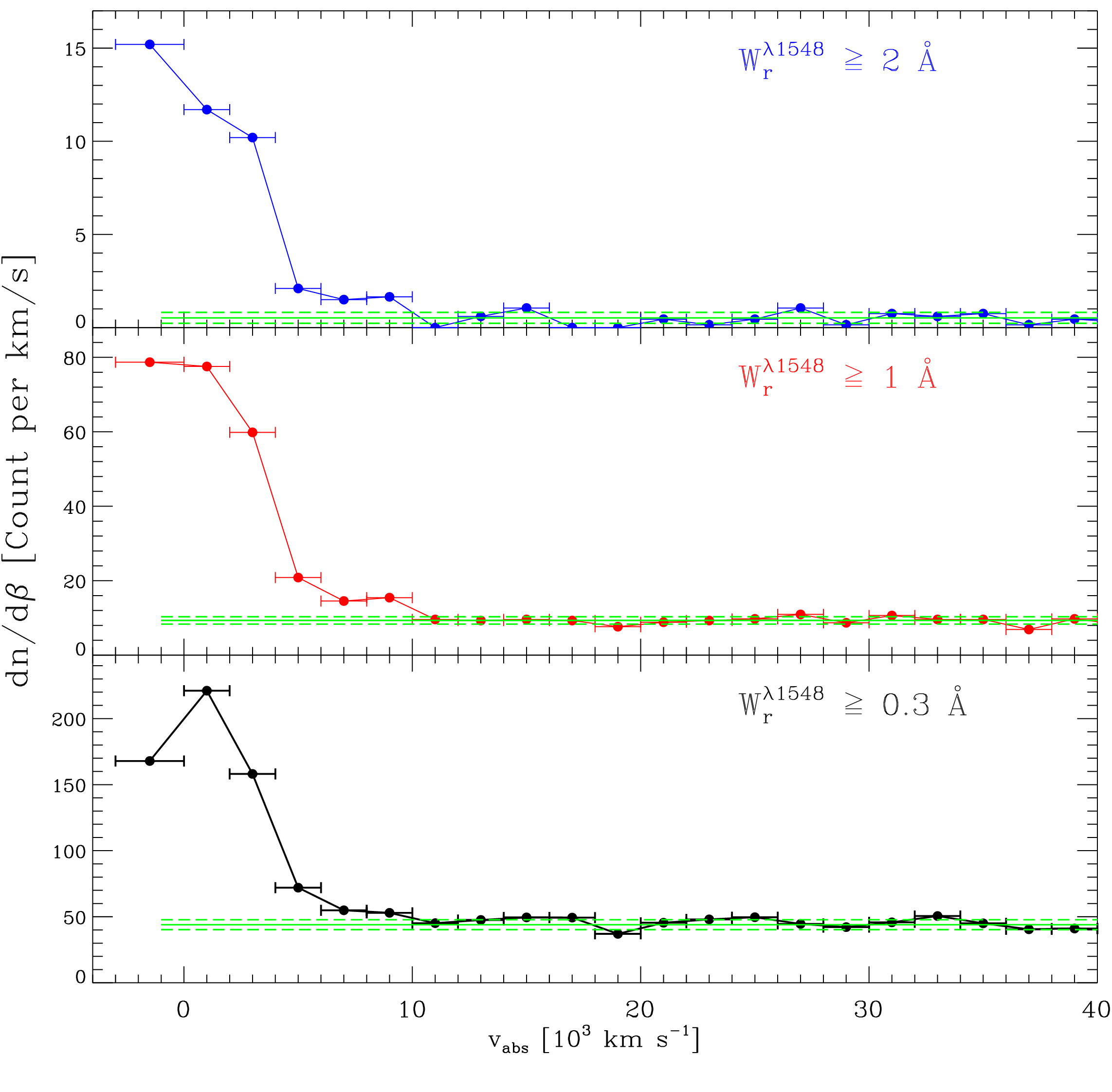}
\caption{The number density of absorbers ($dn/d\beta$) in given velocity interval as a function of velocity offset from the quasars, whose unit is number of absorbers per \kms. The starting \vabs\ is -3000 \kms, and the bin size is 2000 \kms\ for the absorptions with \vabs\ $>0$. The horizonal solid-lines illustrate average values for absorbers far away from the quasars, and the horizonal dash-lines are corresponding $\pm1\sigma$ from Poisson statistics. Left figure is for \MgII\ absorptions, and right one is for \CIV\ absorptions.}
\label{fig:dndbeta}
\end{figure*}

Starting from $v_{\rm abs}=-3000$ \kms\ and using a bin size of 2000 \kms\ for the absorptions with $v_{\rm abs}>0$, we calculate the $dn/d\beta$ and display them as a function of velocity offset in Figure \ref{fig:dndbeta}. These distributions are complex. No matter what the lowest absorption strength limit is, a significantly excessive $dn/d\beta$ at small \vabs\ range is clearly over-plotted on an approximately constant distribution. These distributions tell us that our absorber sample contain absorptions originated in (1) cosmologically intervening structures which are expected to produce uniform $dn/d\beta$; (2) quasar environments including quasar host galaxy, CGM and IGM within the same one cluster/group; (3) quasar outflow/wind. The second contributions are expected to show a normal $dn/d\beta$ distribution that is located at $\beta\approx0$ and significantly higher than a random distribution, and the third contributions would cause an asymmetrically extended blue tail and destroy the normal distribution.

In order to assess the significance level of the $dn/d\beta$ excess at small velocity offsets and make comparisons, we calculate the mean value of $dn/d\beta$ at \vabs\ $>20000$ \kms and corresponding Poisson error. The absorbers with so large \vabs\ are generally regarded as intervening absorptions, though there are some variable NALs with very large velocity separation from the quasar emission redshfit \cite[e.g.,][]{2004ApJ...601..715N,2013MNRAS.434..163H,2011MNRAS.410.1957H,2013MNRAS.434.3275C,2013ApJ...777...56C,2015MNRAS.450.3904C}. The results are illustrated with green horizonal lines in Figure \ref{fig:dndbeta}. For several velocity intervals, we list the significant levels of excessive $dn/d\beta$ with respect to the average $dn/d\beta$ of intervening absorptions in Table \ref{tab:dndbetamg} (\MgII) and Table \ref{tab:dndbetaciv} (\CIV). It is clear that the excessive $dn/d\beta$ at small \vabs\ are very remarkable for both weak and strong absorptions, and that the extensions of the excess are obviously different between \CIV\ and \MgII\ absorptions even if they are limited to the same absorption strength range. The very strong \MgII\ absorptions with $W_r^{\lambda2796}\ge2$ \AA\ show particularly evident excess within 2000 \kms. While the excessive tail of the moderate \MgII\ absorptions with $W_r^{\lambda2796}\ge1$ \AA\ is extended to 4000 \kms\ at a significant level of $4.3\sigma$. In addition, the excessive tail has a significant level of $3\sigma$ at 8000 \kms\ for the weak Mg II absorptions with $W_r^{\lambda2796}\ge0.3$ \AA. These indicate that the excess of \MgII\ absorptions is physically connected to absorption strengths. Turning to \CIV\ absorptions, significant excess is extended to 10000 \kms\ for all of the weak ($W_r^{\lambda1548}\ge0.3$ \AA), moderate ($W_r^{\lambda1548}\ge1$ \AA) and strong ($W_r^{\lambda1548}\ge0.3$ \AA) absorptions. No significant difference among the excessive behaviors of weak, moderate and strong \CIV\ absorptions.

\begin{table}
\caption{The significance level of incidences of \MgII\ absorptions as a function of absorption strength.} \tabcolsep 1.7mm
\centering
\label{tab:dndbetamg}
 \begin{tabular}{lccc}
 \hline\hline\noalign{\smallskip}
 \multirow{2}{*}{velocity \kms}& \multicolumn{3}{c}{Significant level}\\
\cline{2-4}\noalign{\smallskip}
&  $W_r^{\lambda2796}\ge0.3$ \AA &$W_r^{\lambda2796}\ge1$ \AA & $W_r^{\lambda2796}\ge2$ \AA \\
\hline\noalign{\smallskip}
-3000$<$\vabs$<$0   & 13.2 & 10.2 & 6.2 \\
0$<$\vabs$<$2000    & 44.4 & 43.8 & 25.0 \\
2000$<$\vabs$<$4000 & 5.8  & 4.3  & 1.8 \\
4000$<$\vabs$<$6000 & 1.4  & 1.0 & 0.6\\
6000$<$\vabs$<$8000 & 3.0  & 2.3 & 0.9\\
8000$<$\vabs$<$10000& 1.0  & 1.5 & 1.3\\
\noalign{\smallskip}
\hline\hline\noalign{\smallskip}
\end{tabular}
\begin{flushleft}
Note --- Significant levels of excess relative to the random occurrence of intervening absorptions.
\end{flushleft}
\end{table}

\begin{table}
\caption{The significance level of incidences of \CIV\ absorptions as a function of absorption strength.} \tabcolsep 1.7mm
\centering
\label{tab:dndbetaciv}
 \begin{tabular}{lccc}
 \hline\hline\noalign{\smallskip}
 \multirow{2}{*}{velocity \kms}& \multicolumn{3}{c}{Significant level}\\
\cline{2-4}\noalign{\smallskip}
&  $W_r^{\lambda1548}\ge0.3$ \AA &$W_r^{\lambda1548}\ge1$ \AA & $W_r^{\lambda1548}\ge2$ \AA \\
\hline\noalign{\smallskip}
-3000$<$\vabs$<$0    & 33.0 & 70.6 & 50.0 \\
0$<$\vabs$<$2000     & 47.2 & 69.4 & 38.0 \\
2000$<$\vabs$<$4000  & 30.4 & 51.4 & 32.9 \\
4000$<$\vabs$<$6000  & 7.5 & 11.7 & 5.4 \\
6000$<$\vabs$<$8000  & 2.9 & 5.3 & 3.3 \\
8000$<$\vabs$<$10000 & 2.4 & 6.2  & 3.8 \\
10000$<$\vabs$<$12000& 0.3 & 0.3  & -1.8 \\
\noalign{\smallskip}
\hline\hline\noalign{\smallskip}
\end{tabular}
\begin{flushleft}
Note --- Significance levels of excess relative to the random occurrence of intervening absorptions.
\end{flushleft}
\end{table}

\begin{figure*}
\centering
\includegraphics[width=0.47\textwidth]{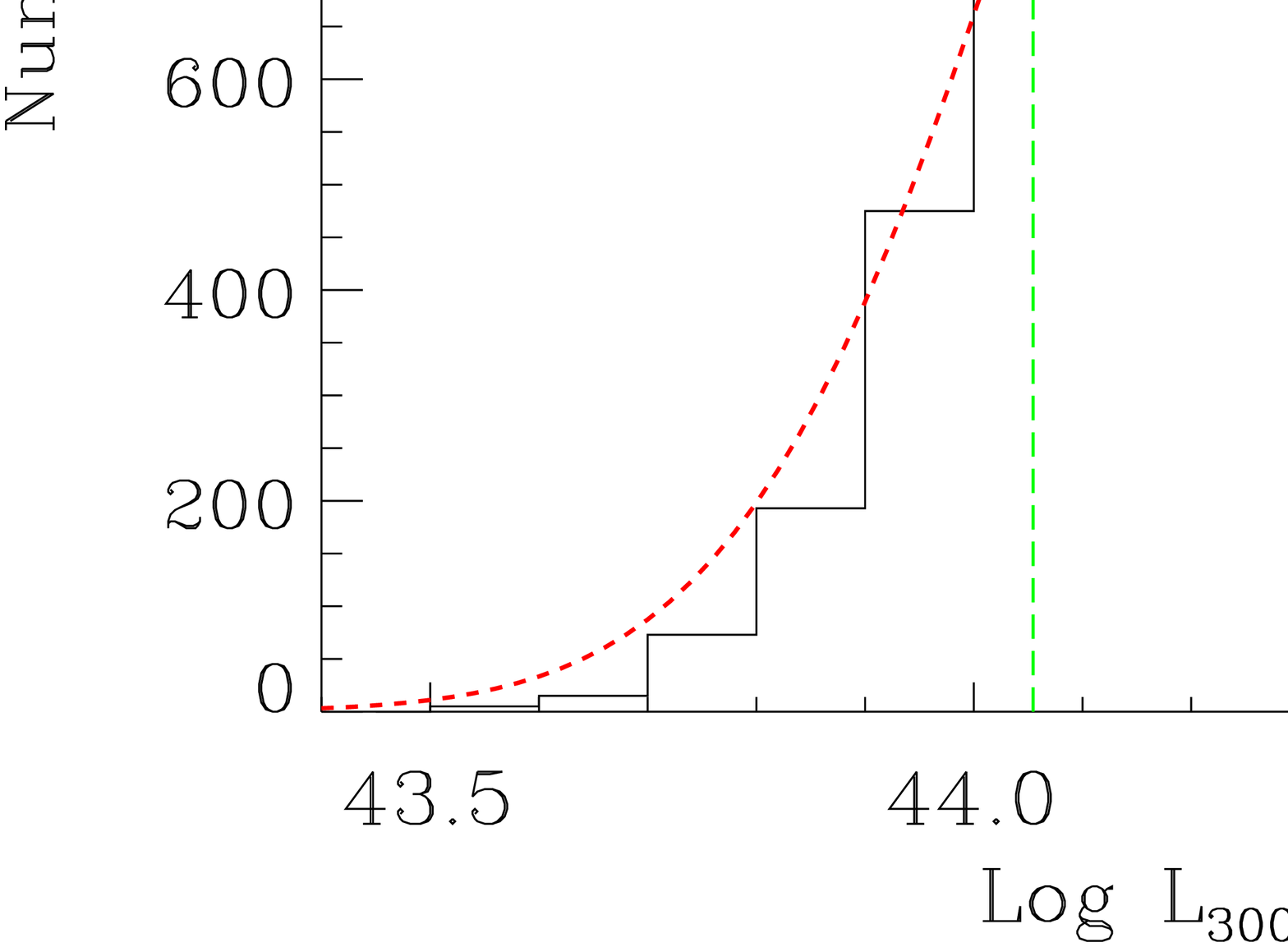}
\hspace{3ex}
\includegraphics[width=0.47\textwidth]{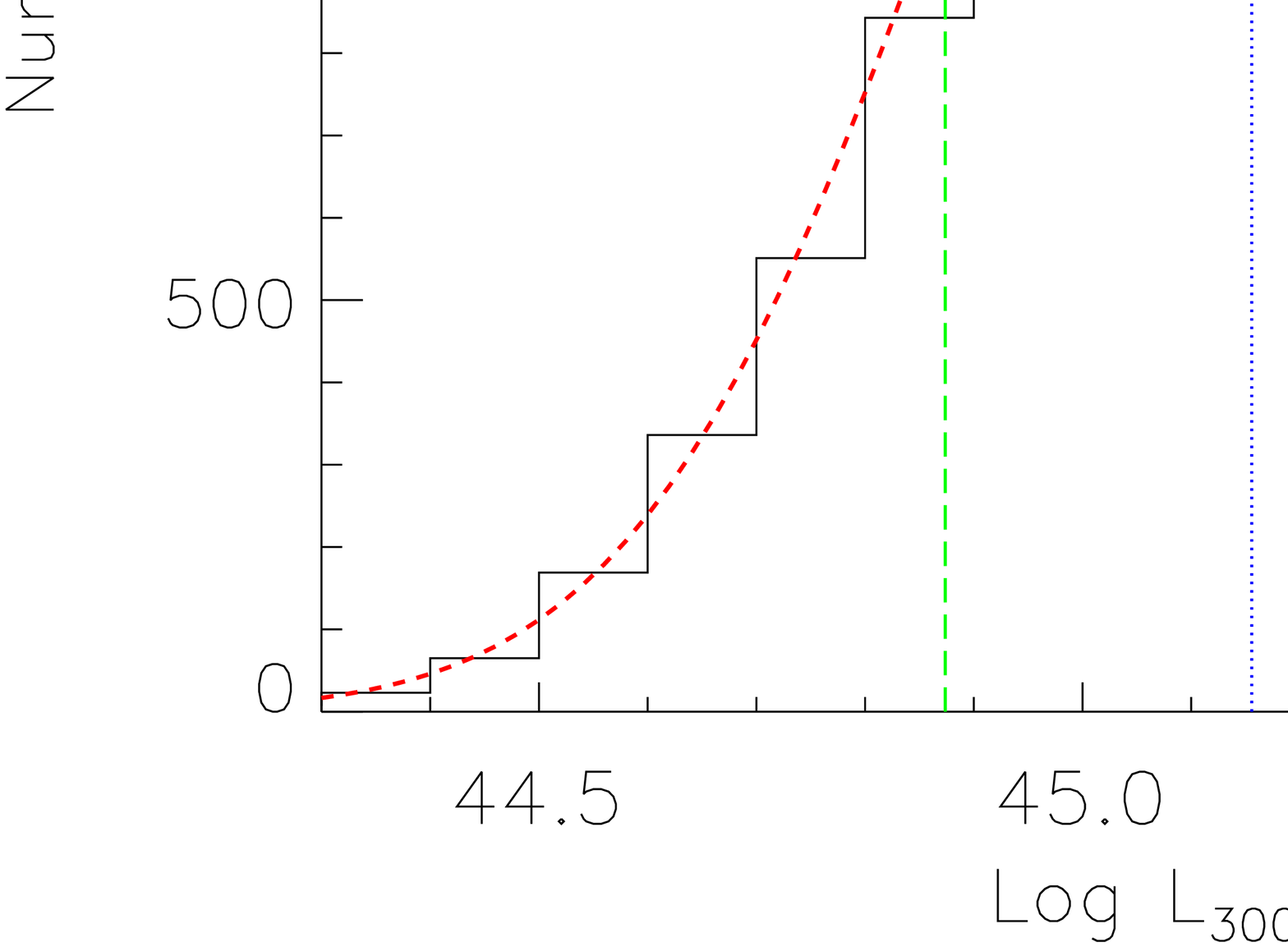}
\caption{Continuum luminosities of quasars at rest-frame 3000 \AA. Red dash-curves represent the Gaussian fittings. Blue dash-lines label the centers of Gaussian fittings ($L_{\rm center}$), and green dash-lines illustrate the positions deviated from the Gaussian centers $\pm\sigma$, where $\sigma$ are the standard deviations produced by the Gaussian fittings.}
\label{fig:L3000_distr}
\end{figure*}

\begin{figure*}
\centering
\includegraphics[width=0.47\textwidth]{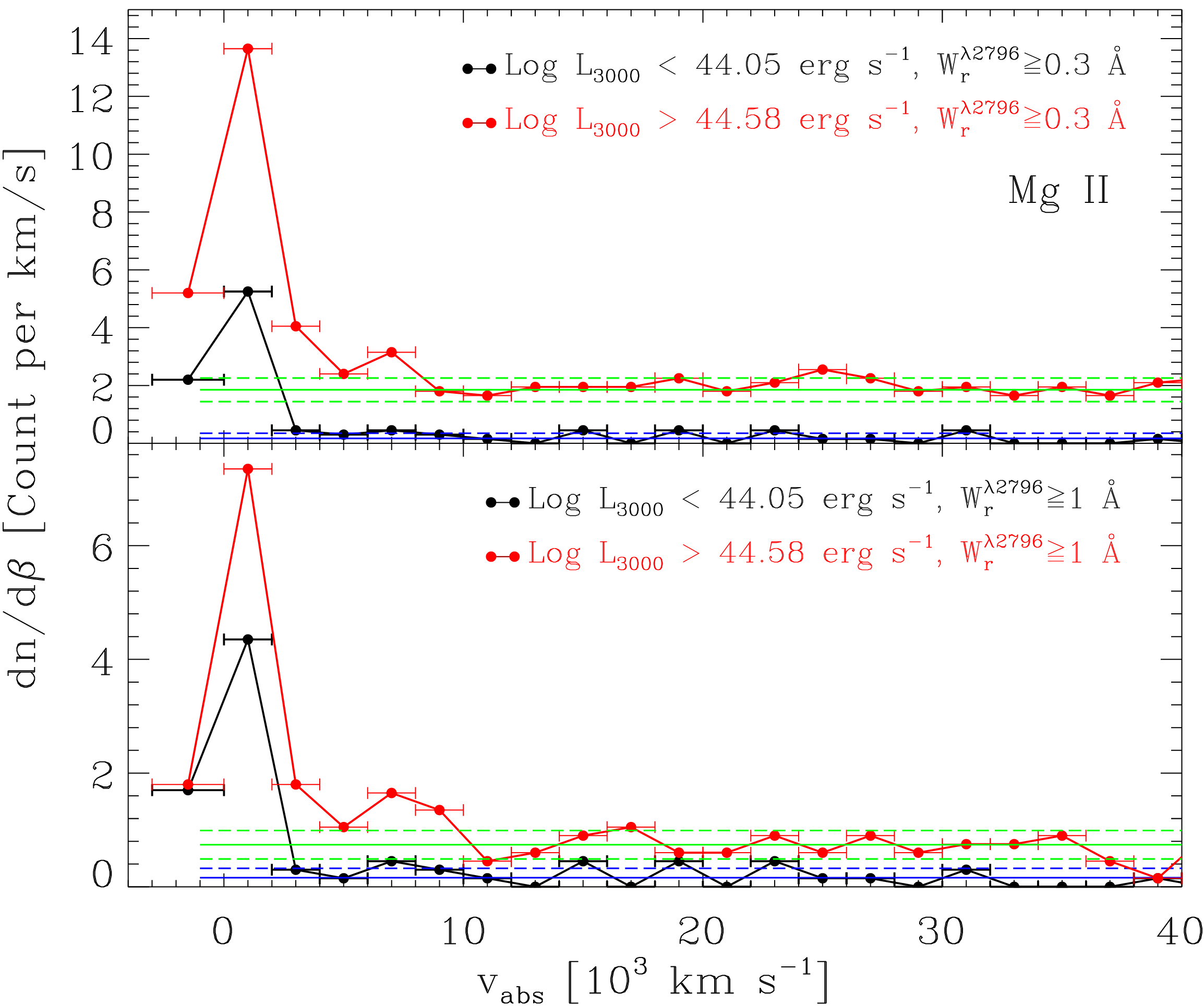}
\hspace{3ex}
\includegraphics[width=0.47\textwidth]{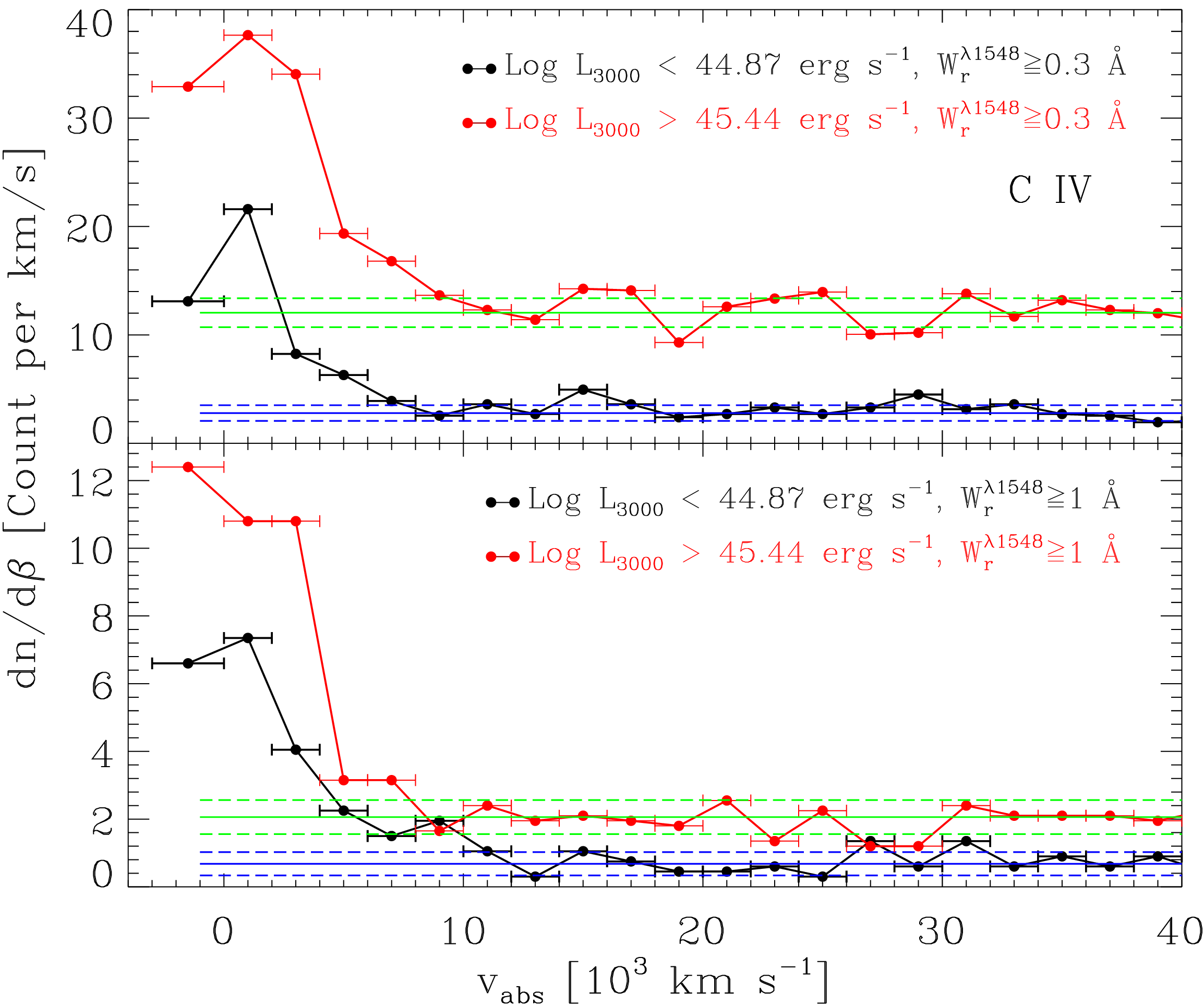}
\caption{Number density of absorbers in a given velocity interval as a function of velocity offset from the quasars ($dn/d\beta$). The starting \vabs\ is -3000 \kms, and the bin size is 2000 \kms\ for the absorptions with \vabs\ $>0$. The horizonal solid lines illustrate average values for absorbers far away from the quasars, and the horizonal dash lines are corresponding $\pm1\sigma$ from Poisson statistics. Black symbols represent the low-luminosity quasar sample, and red symbols represent the high-luminosity quasar sample. Left figure is for \MgII\ absorptions, and right one is for \CIV\ absorptions.}
\label{fig:dndbeta_L}
\end{figure*}

It is widely accepted that the quasar radiation can affect a quasar's surrounding environment. Here, we investigate the possible dependence of the $dn/d\beta$ on the quasar luminosity (continuum luminosity can be converted to bolometric luminosity by multiplying by a factor). We calculate quasar continuum luminosity at rest-frame 3000 \AA\ ($L_{\rm3000}$) by directly measuring the pseudo-continuum fitting of the quasar spectra \cite[e.g.,][]{2015ApJS..221...32C,2016MNRAS.462.2980C}. The results are displayed in Figure \ref{fig:L3000_distr}. In Figure \ref{fig:L3000_distr}, the fittings of Gaussian function are over-plotted with red dash curves, the centers ($L_{\rm center}$) of the Gaussian fittings are marked with vertical blue dash lines, and $\pm1\sigma$ deviations from the centers are labeled with vertical green dash lines, where the $\sigma$ is the standard deviation returned by the Gaussian fitting. Here we construct low- and high-luminosity quasar samples with $L_{\rm3000}<L_{\rm center}-\sigma$ and $L_{\rm3000}>L_{\rm center}+\sigma$, respectively. For \MgII\ absorptions, low- and high-luminosity quasars have $L_{\rm3000}<10^{44.05}$ \ergs\ and $L_{\rm3000}>10^{44.58}$ \ergs, respectively. For \CIV\ absorptions, low- and high-luminosity quasars have $L_{\rm3000}<10^{44.87}$ \ergs\ and $L_{\rm3000}>10^{45.44}$ \ergs, respectively. For the low- and high-luminosity quasar samples, the number densities of absorbers are displayed in Figure \ref{fig:dndbeta_L}, and the significance levels of excessive $dn/d\beta$ are provided in Table \ref{tab:dndbeta_Lmg} (\MgII) and Table \ref{tab:dndbeta_Lciv} (\CIV).

\begin{table}
\caption{The significance level of incidences of \MgII\ absorptions as a function of quasar luminosity.} \tabcolsep 1mm
\centering
\label{tab:dndbeta_Lmg}
 \begin{tabular}{lccccc}
 \hline\hline\noalign{\smallskip}
 \multirow{3}{*}{velocity \kms}& \multicolumn{4}{c}{Significant level}\\
\cline{2-5}\noalign{\smallskip}
&  \multicolumn{2}{c}{$L_{\rm3000}<10^{44.05}$ \ergs} & \multicolumn{2}{c}{$L_{\rm3000}>10^{44.58}$ \ergs} \\
\cline{2-5}\noalign{\smallskip}
& \tiny{$W_r^{\lambda2796}\ge0.3$ \AA} & \tiny{$W_r^{\lambda2796}\ge1$ \AA} &  \tiny{$W_r^{\lambda2796}\ge0.3$ \AA} & \tiny{$W_r^{\lambda2796}\ge1$ \AA} \\
\hline\noalign{\smallskip}
-3000$<$\vabs$<$0   & 11.4 & 9.3  & 8.2  & 4.2 \\
0$<$\vabs$<$2000    & 28.5 & 25.4 & 29.0 & 26.4 \\
2000$<$\vabs$<$4000 & 1.6  & 0.8  & 5.4  & 4.2 \\
4000$<$\vabs$<$6000 & 0.7  & 0.0  & 1.3  & 1.2 \\
6000$<$\vabs$<$8000 & 1.6  & 1.8  & 3.2  & 3.6 \\
8000$<$\vabs$<$10000& 0.7  & 0.8  & -0.1 & 2.4 \\
\noalign{\smallskip}
\hline\hline\noalign{\smallskip}
\end{tabular}
\begin{flushleft}
Note --- Significant levels of excess relative to the random occurrence of intervening absorptions.
\end{flushleft}
\end{table}

\begin{table}
\caption{The significant level of incidences of \CIV\ absorptions as a function of quasar luminosity.} \tabcolsep 1mm
\centering
\label{tab:dndbeta_Lciv}
 \begin{tabular}{lccccc}
 \hline\hline\noalign{\smallskip}
 \multirow{3}{*}{velocity \kms}& \multicolumn{4}{c}{Significant level}\\
\cline{2-5}\noalign{\smallskip}
&  \multicolumn{2}{c}{$L_{\rm3000}<10^{44.87}$ \ergs} & \multicolumn{2}{c}{$L_{\rm3000}>10^{45.44}$ \ergs} \\
\cline{2-5}\noalign{\smallskip}
& \tiny{$W_r^{\lambda1548}\ge0.3$ \AA} & \tiny{$W_r^{\lambda1548}\ge1$ \AA} &  \tiny{$W_r^{\lambda1548}\ge0.3$ \AA} & \tiny{$W_r^{\lambda1548}\ge1$ \AA} \\
\hline\noalign{\smallskip}
-3000$<$\vabs$<$0   & 14.3 & 17.2 & 15.6 & 20.6 \\
0$<$\vabs$<$2000    & 26.0 & 19.4 & 19.2 & 17.4 \\
2000$<$\vabs$<$4000 & 7.5  & 9.8  & 16.5 & 17.4 \\
4000$<$\vabs$<$6000 & 4.9  & 4.6  & 5.5  & 2.2 \\
6000$<$\vabs$<$8000 & 1.5  & 2.4  & 3.6  & 2.2 \\
8000$<$\vabs$<$10000& -0.3 & 3.7  & 1.2  & -0.8 \\
10000$<$\vabs$<$12000& 1.1 & 1.1  & 0.2  & 0.6 \\
\noalign{\smallskip}
\hline\hline\noalign{\smallskip}
\end{tabular}
\begin{flushleft}
Note --- Significant levels of excess relative to the random occurrence of intervening absorptions.
\end{flushleft}
\end{table}

We note from Figure \ref{fig:dndbeta_L} and Table \ref{tab:dndbeta_Lmg} that the excessive $dn/d\beta$ of \MgII\ absorptions with $W_r^{\lambda2796}\ge0.3$ \AA\ of high-luminosity quasars could be extended up to 8000 \kms, while that of low-luminosity quasars is mainly limited within 2000 \kms. These results are similar to those of \cite{2013JApA...34..357P}. That is, the luminous quasars exhibits a longer extension of excessive $dn/d\beta$ when compared to the faint quasars. Turning to the \CIV\ absorptions with $W_r^{\lambda1548}\ge0.3$ \AA\ , no obvious difference between the $dn/d\beta$ extensions of faint and luminous quasars is seen from Figure \ref{fig:dndbeta_L} and Table \ref{tab:dndbeta_Lciv}, which is consistent with the findings of \cite{2016MNRAS.462.3285P}.

\begin{figure*}
\centering
\includegraphics[width=0.47\textwidth]{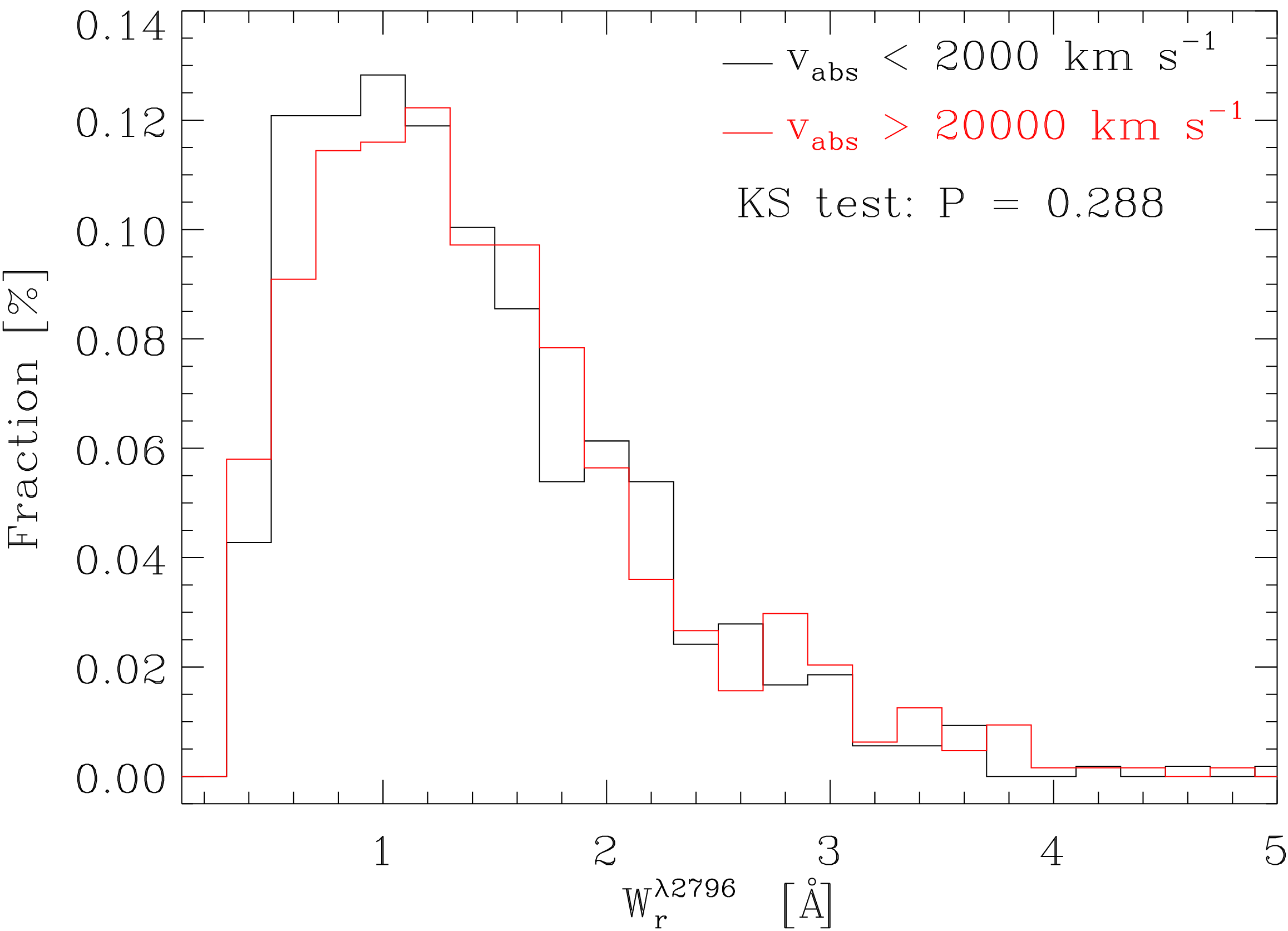}
\hspace{3ex}
\includegraphics[width=0.47\textwidth]{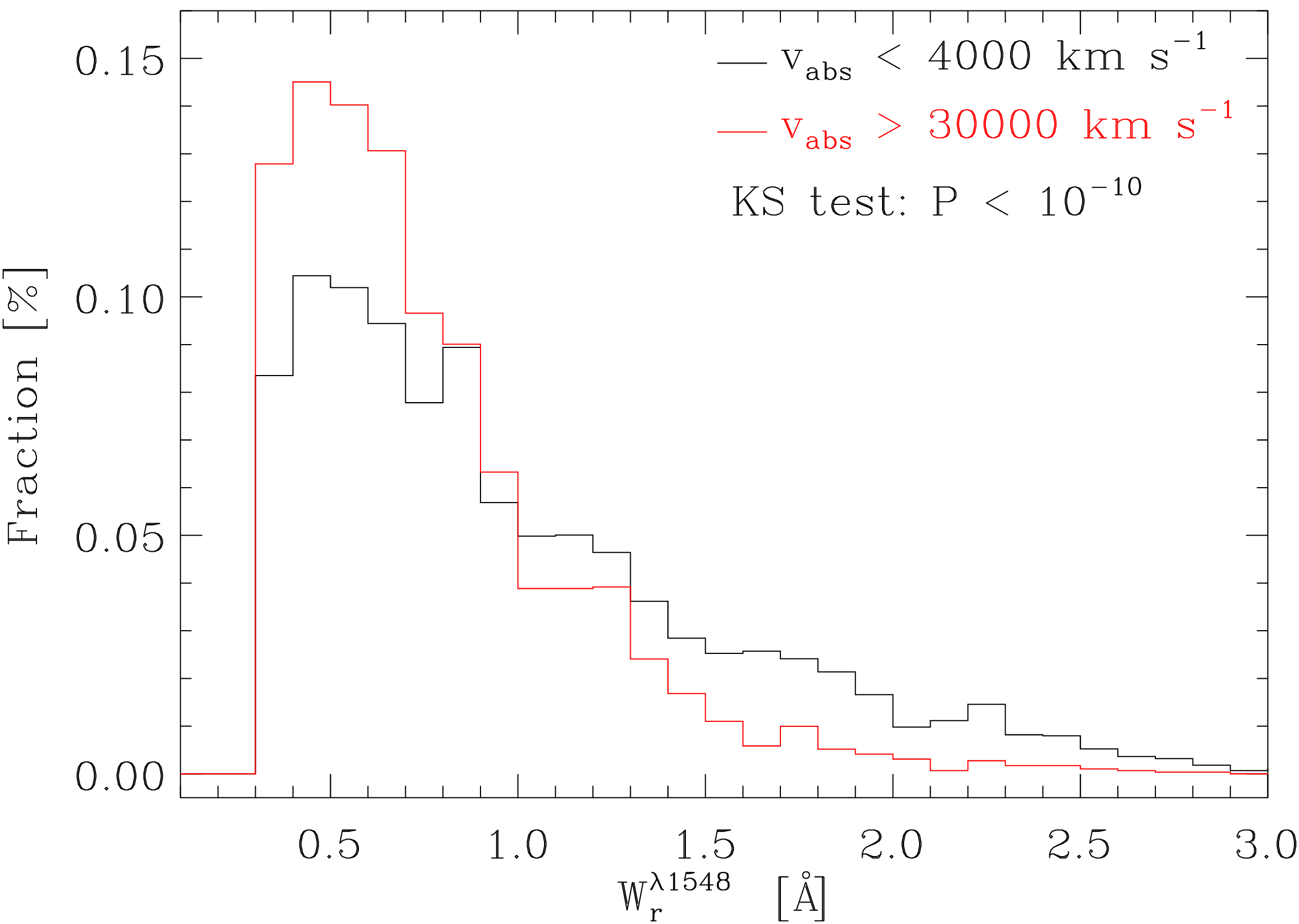}
\caption{Distributions of equivalent widths at rest-frame. Left panel is for \MgII\ absorptions, and right one is for \CIV\ absorptions. Black lines represent the associated absorptions, and red lines represent the intervening ones. Y-axis indicates the number of absorbers that have been normalized by the total number of absorbers within each subsample.}
\label{fig:ew}
\end{figure*}

We note that the luminous quasars on average have higher continuum signal-to-noise ratio than the faint quasars, hence will have more detectable absorbers. We explore the possibility that luminosity dependence of the $dn/d\beta$ is due to the detection probability of absorbers. Figure \ref{fig:ew} shows the equivalent widths at rest-frame for the absorptions with low and high velocity offsets from the quasars. The Kolmogorov-Smirnov (KS) test suggests that \MgII\ absorptions with low velocity offset have an equivalent width distribution similar to that of intervening \MgII\ absorptions, which indicates that both the associated and intervening \MgII\ absorptions would have similar detection probability. In addition, Figure \ref{fig:dndbeta_L} and Table \ref{tab:dndbeta_Lmg} clearly show that the \MgII\ absorptions with $W_r^{\lambda2796}\ge1$ \AA\ also exhibit a different extension of excessive $dn/d\beta$ for the low- and high-luminosity quasars. The detection of stronger absorptions are expected to be less affected by the different quasar luminosity. Therefore, there is no obvious correlation between the incidence of \MgII\ absorbers and the absorber detection probability. And thus, the difference of the \MgII\ $dn/d\beta$ distributions between low- and high-luminosity quasars is not originated from absorber detection probability.

Turning to the \CIV\ absorptions, the KS test suggests that the equivalent width distribution of the \CIV\ absorptions with low velocity offset is different from that of the intervening ones. The \CIV\ absorptions with low velocity offset on average have a larger equivalent width than the intervening ones, which suggests that \CIV\ absorptions with low velocity offset would have a slightly high detection probability relative to the intervening ones. While, the $dn/d\beta$ and $f_c$ (see Section \ref{sect:fc_wr} and Figure \ref{fig:cf_w}) distributions do not show an obvious dependence on absorption strength relative to the background intervening absorptions. In addition, Figure \ref{fig:dndbeta_L} and Table \ref{tab:dndbeta_Lmg} also indicate that, either \CIV\ absorptions with $W_r^{\lambda1548}\ge0.3$ \AA\ or $\ge1$ \AA, there in no evident difference of relative excess $dn/d\beta$ between low and high luminosity quasars. The slightly different absorption strength dependences of the relative excess $dn/d\beta$ can be ascribed to the slightly higher equivalent width of the associated absorptions.

\begin{figure}
\centering
\includegraphics[width=0.47\textwidth]{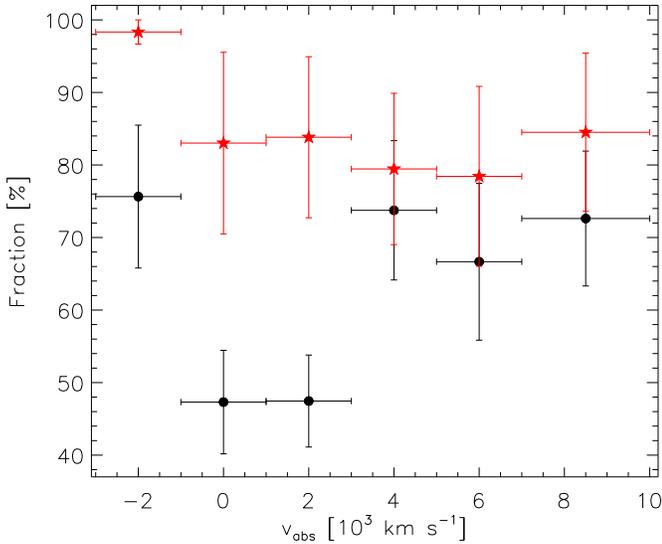}
\caption{Fraction of \CIV\ (\MgII) absorbers that have a \MgII\ (\CIV) absorption counterparts as a function of absorber velocity offset from the quasar. The black circles are for \CIV\ absorbers and the red stars are for \MgII\ absorbers. The $\pm1\sigma$ errors of the fractions are from Poisson statistics.}
\label{fig:vabsciv_mg}
\end{figure}

The \CIV\ quasar sample covers both the \CIV\ and \MgII\ absorption doublets. Some \CIV\ associated absorptions might have \MgII\ absorption counterparts, and vice versa. We directly adopt the \MgII\ absorption systems with $v_{\rm abs}<10000$ \kms included in the absorption catalog of \cite{2015ApJS..221...32C} to search for the counterparts of the \CIV\ absorptions with $v_{\rm abs}<10000$ \kms. We find that there are 238 quasars whose spectra emerge simultaneously detectable \CIV\ and \MgII\ absorptions with $v_{\rm abs}<10000$ \kms. In these 238 quasar spectra, there are 510 \CIV\ and 376 \MgII\ absorption systems with $v_{\rm abs}<10000$ \kms. In several velocity intervals, we estimate the fraction of \CIV\ absorbers that have a \MgII\ absorption counterpart with a velocity uncertainty of 200 \kms, and vice versa. The results are displayed in Figure \ref{fig:vabsciv_mg}. It is clear that most of the \MgII\ absorbers have a \CIV\ absorption counterpart, and the fraction of the \MgII\ absorbers is significantly higher than that of the \CIV\ absorbers when the absorbers are limited within $v_{\rm abs}<3000$ \kms.

\subsection{Coverage fraction and absorption strength}
\label{sect:fc_wr}
\begin{figure*}
\centering
\includegraphics[width=0.47\textwidth]{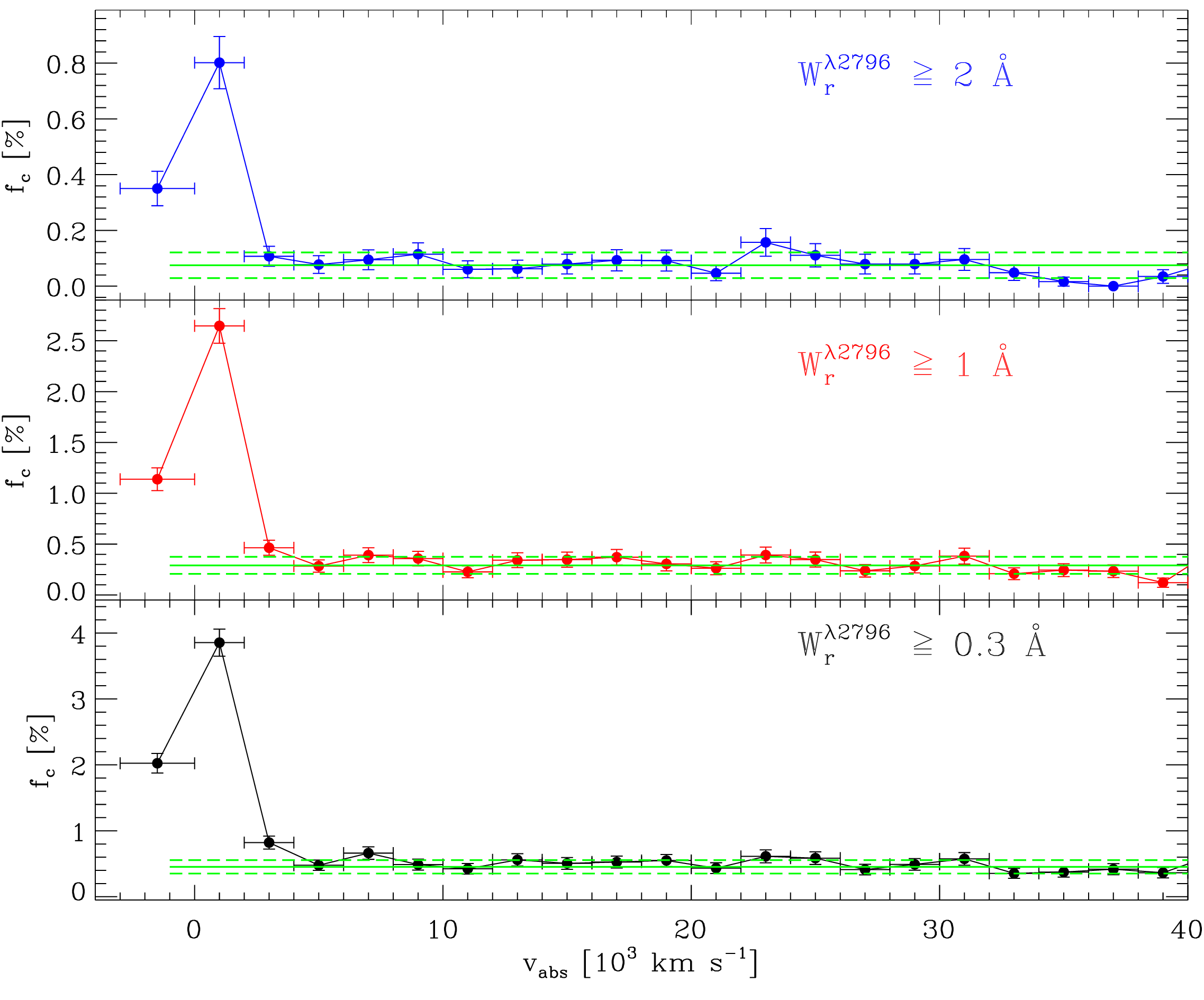}
\hspace{3ex}
\includegraphics[width=0.47\textwidth]{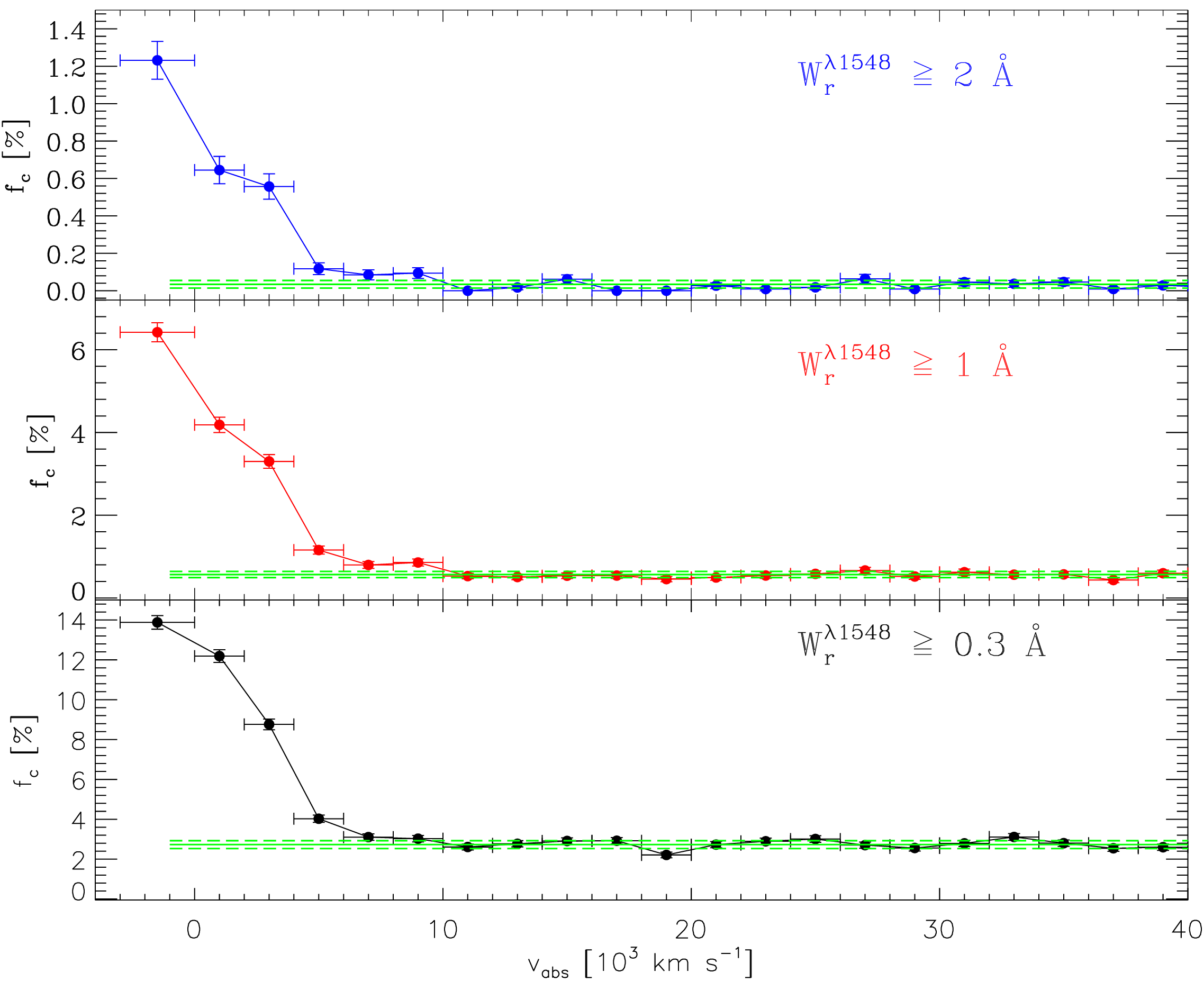}
\caption{The coverage fractions of absorptions ($f_c$), which are the ratio between the number of  quasars with at least one detected absorber and the total number of quasars can be used to search for corresponding absorptions in given bins of velocity offset from quasar redshifts. The starting \vabs\ is -3000 \kms, and the bin size is 2000 \kms\ for the absorptions with \vabs\ $>0$. The horizonal solid-lines illustrate average fractions for absorbers far away from the quasars, and the horizonal dash-lines are corresponding $\pm1\sigma$ from Poisson statistics. Left figure is for \MgII\ absorptions, and right one is for \CIV\ absorptions.}
\label{fig:cf_w}
\end{figure*}

We define the coverage fraction of absorptions, $f_c$, as the ratio of the number of quasars with at least one detected absorber to the total number of quasars that can be used to detect corresponding absorptions, within a given bin of velocity offset from the quasar system. The error of $f_c$ can be estimated from Poisson statistics. One would expect that the $f_c$ is related to absorption strength. Hence, here we measure the $f_c$ in several ranges of absorption strengths as a function of velocity offset from the quasar system. The measurements are provided in Figure \ref{fig:cf_w}. For both \CIV\ and \MgII\ absorptions, no matter what the lowest limit of absorption strength is, we can observe an approximately constant $f_c$ at large velocity offset, which is expected for the cosmologically intervening absorptions. For comparisons, we calculate the mean value of $f_c$ at \vabs\ $>20000$ \kms and corresponding Poisson error. The results are illustrated with green horizonal lines in Figure \ref{fig:cf_w}, and the significant levels of excessive $f_c$ with respect to the random occurrence of intervening absorptions are listed in Table \ref{tab:fcmg} (\MgII) and Table \ref{tab:fcciv} (\CIV).

\begin{table}
\caption{The significant level of coverage fraction of \MgII\ absorptions as a function of absorption strength.} \tabcolsep 1.7mm
\centering
\label{tab:fcmg}
 \begin{tabular}{lccc}
 \hline\hline\noalign{\smallskip}
 \multirow{2}{*}{velocity \kms}& \multicolumn{3}{c}{Significant level}\\
\cline{2-4}\noalign{\smallskip}
&  $W_r^{\lambda2796}\ge0.3$ \AA &$W_r^{\lambda2796}\ge1$ \AA & $W_r^{\lambda2796}\ge2$ \AA \\
\hline\noalign{\smallskip}
-3000$<$\vabs$<$0   & 15.4 & 10.1  & 6.0 \\
0$<$\vabs$<$2000    & 33.4 & 28.1 & 15.8 \\
2000$<$\vabs$<$4000 & 3.6  & 2.1  & 0.7 \\
4000$<$\vabs$<$6000 & 0.2  & 0.0  & 0.1\\
6000$<$\vabs$<$8000 & 2.1  & 1.2  & 0.4\\
8000$<$\vabs$<$10000& 0.3  & 0.8  & 0.9\\
\noalign{\smallskip}
\hline\hline\noalign{\smallskip}
\end{tabular}
\begin{flushleft}
Note --- Significant levels of excess relative to the random occurrence of intervening absorptions.
\end{flushleft}
\end{table}

\begin{table}
\caption{The significant level of coverage fraction of \CIV\ absorptions as a function of absorption strength.} \tabcolsep 1.7mm
\centering
\label{tab:fcciv}
 \begin{tabular}{lccc}
 \hline\hline\noalign{\smallskip}
 \multirow{2}{*}{velocity \kms}& \multicolumn{3}{c}{Significant level}\\
\cline{2-4}\noalign{\smallskip}
&  $W_r^{\lambda1548}\ge0.3$ \AA &$W_r^{\lambda1548}\ge1$ \AA & $W_r^{\lambda1548}\ge2$ \AA \\
\hline\noalign{\smallskip}
-3000$<$\vabs$<$0   & 56.3 & 78.3 & 59.3 \\
0$<$\vabs$<$2000    & 47.7 & 48.4 & 30.3 \\
2000$<$\vabs$<$4000 & 30.4 & 36.6 & 25.9 \\
4000$<$\vabs$<$6000 & 6.5  & 7.9  & 4.1 \\
6000$<$\vabs$<$8000 & 1.9  & 3.1  & 2.5 \\
8000$<$\vabs$<$10000& 1.5  & 3.9  & 3.0 \\
10000$<$\vabs$<$12000& -0.6 & -0.5 & -1.6 \\
\noalign{\smallskip}
\hline\hline\noalign{\smallskip}
\end{tabular}
\begin{flushleft}
Note --- Significant levels of excess relative to the random occurrence of intervening absorptions.
\end{flushleft}
\end{table}

The distribution behaviors of the $f_c$ are similar to those of the $dn/d\beta$ (see Figure \ref{fig:dndbeta}). The excessive $f_c$ at small \vabs\ are very obvious relative to the constant $f_c$. No matter what the \CIV\ absorption strength is, the obvious excess is extended to 6000 \kms\ at a high significance level of $>4\sigma$ and could be sustained beyond 10000 \kms. This is consistent with the results of \cite{2016MNRAS.462.3285P}. The excessive $f_c$ extensions of \MgII\ absorptions are seemingly related to absorption strengths. The $f_c$ excess of the very strong \MgII\ absorptions with $W_r^{\lambda2796}\ge2$ \AA\ is constrained within 2000 \kms, whereas the weak \MgII\ absorptions with $W_r^{\lambda2796}\ge0.3$ \AA\ could hold excessive $f_c$ beyond 8000 \kms. In addition, in the same velocity offset bin, the weaker \MgII\ absorptions exhibit the more significant excess.

Figure \ref{fig:cf_w} and Tables \ref{tab:fcmg} and \ref{tab:fcciv} clearly indicate that the high excesses of the $f_c$ within 2000 \kms\ for \MgII\ absorptions or within 4000 \kms\ for \CIV\ absorptions dramatically decline to much lower values at higher velocity offsets. The quasar environment absorptions of the host galaxies, CGMs, and IGM within the quasar host galaxy cluster/group, and the low velocity quasar outflow/wind absorptions would dominate the significantly excessive $f_c$ within 2000 \kms\ or 4000 \kms. The high velocity outflow absorptions would be the principal contributions for the evidently excessive $f_c$ beyond 2000 \kms\ for \MgII\ absorptions and 4000 \kms\ for \CIV\ absorptions.

Figure \ref{fig:cf_w} clearly shows that the excessive tail of \CIV\ absorptions could be extended beyond 10000 \kms, which is much farther than that of \MgII\ absorptions. This implies that the maximum velocity of \CIV\ outflows is much larger than that of \MgII\ outflows. In addition, Tables \ref{tab:fcmg} and \ref{tab:fcciv} definitely tell us that the excess of \MgII\ absorptions is much less significant than that of \CIV\ absorptions, especially for the excess beyond 2000 \kms.  These different behaviors between \CIV\ and \MgII\ absorptions are possibly related to quasar radiation and ionization potentials of $\rm Mg^+$ and $\rm C^{3+}$ ionized gas. The $\rm Mg^+$ and $\rm C^{3+}$ ions have ionization potentials of 15.035 eV and 64.49 eV, respectively, which indicate that the \CIV\ absorbing clouds can be remained but the \MgII\ ones would have been destroyed when the energy of the incident photons is up to 64.49 eV. One generally expects that quasar outflow hosts very strong radiation field and highly ionizes gas within it. Therefore, quasar outflow would be a greenhouse for \CIV\ absorbing gas but fatal for \MgII\ one. This would be an important reason why the quasars with \CIV\ BALs, which are generally believed to be formed in outflows, are much more than those with \MgII\ BALs. Central gravitation, radiation pressure, thermal pressure, and magnetocentrifugal fores would be the most important mechanisms that drive the quasar outflow. One expects that quasar outflow has been accelerated until the central gravitation is more significant than a combination of another three mechanisms. In addition, we expect that radiation field of quasar outflow in accelerated phase would be much stronger than that in decelerated phase. In this case, intense quasar radiation would result in most of \MgII\ outflows that are in decelerated phase. Therefore, very few \MgII\ outflows exhibit high velocity and most of them host low velocity. In addition, gas clouds with low ionization potential and high column density would be partly ionized by intense quasar radiation and then become the low column density ones, so that stronger \MgII\ absorptions with less significant excess (see Figure \ref{fig:cf_w} and Table \ref{tab:fcmg}). While, high ionization potential of $\rm C^{3+}$ ion lets significant fraction of \CIV\ absorbing gas be sustained in the accelerated phase of quasar outflow. These would be reasonable explanations why the excessive $f_c$ of \CIV\ absorptions is much more obvious than that of \MgII\ absorptions, especially for those with high velocity.

\subsection{Coverage fraction and quasar luminosity}
\label{sect:fc_L}
\begin{figure*}
\centering
\includegraphics[width=0.47\textwidth]{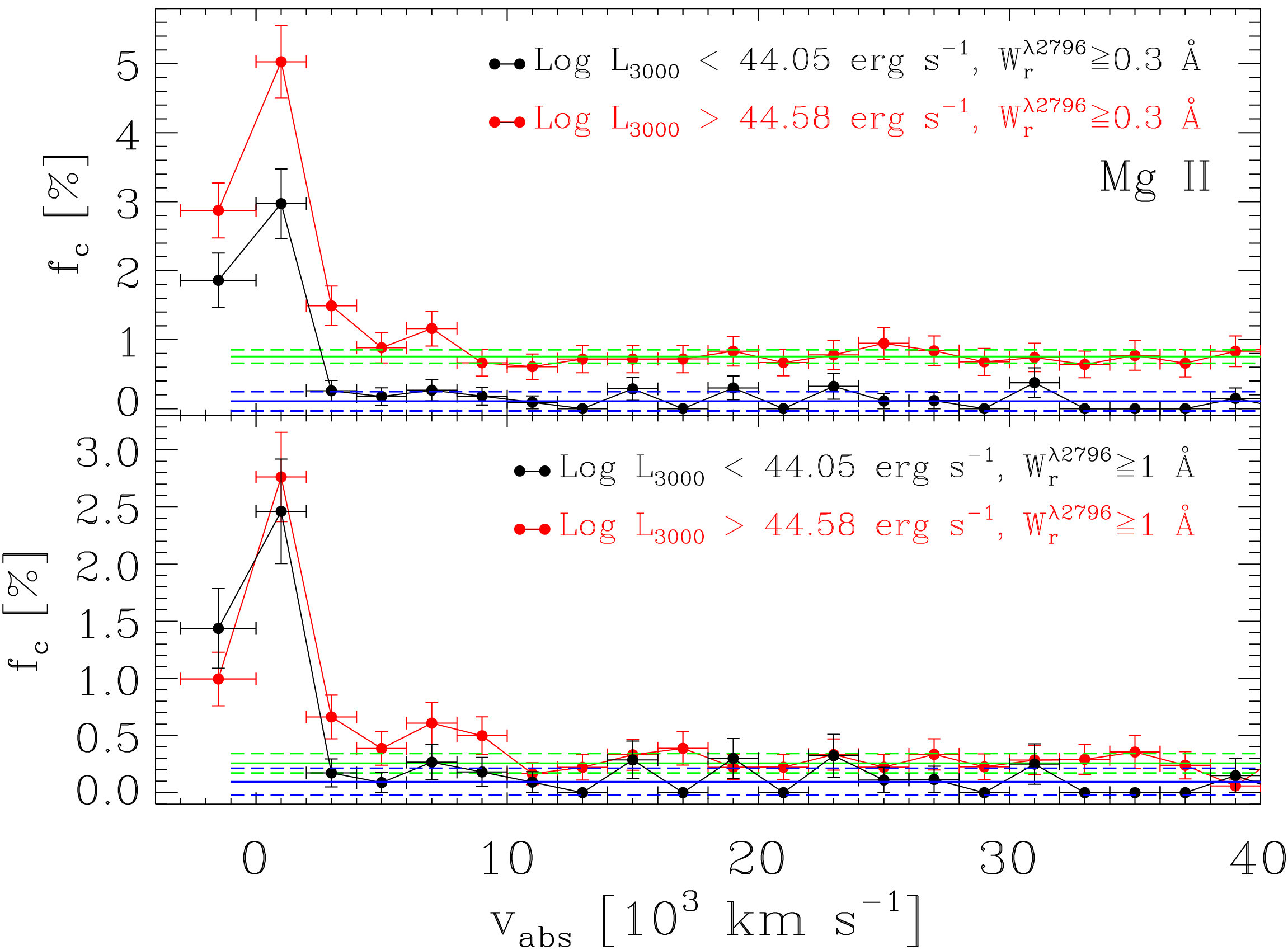}
\hspace{3ex}
\includegraphics[width=0.47\textwidth]{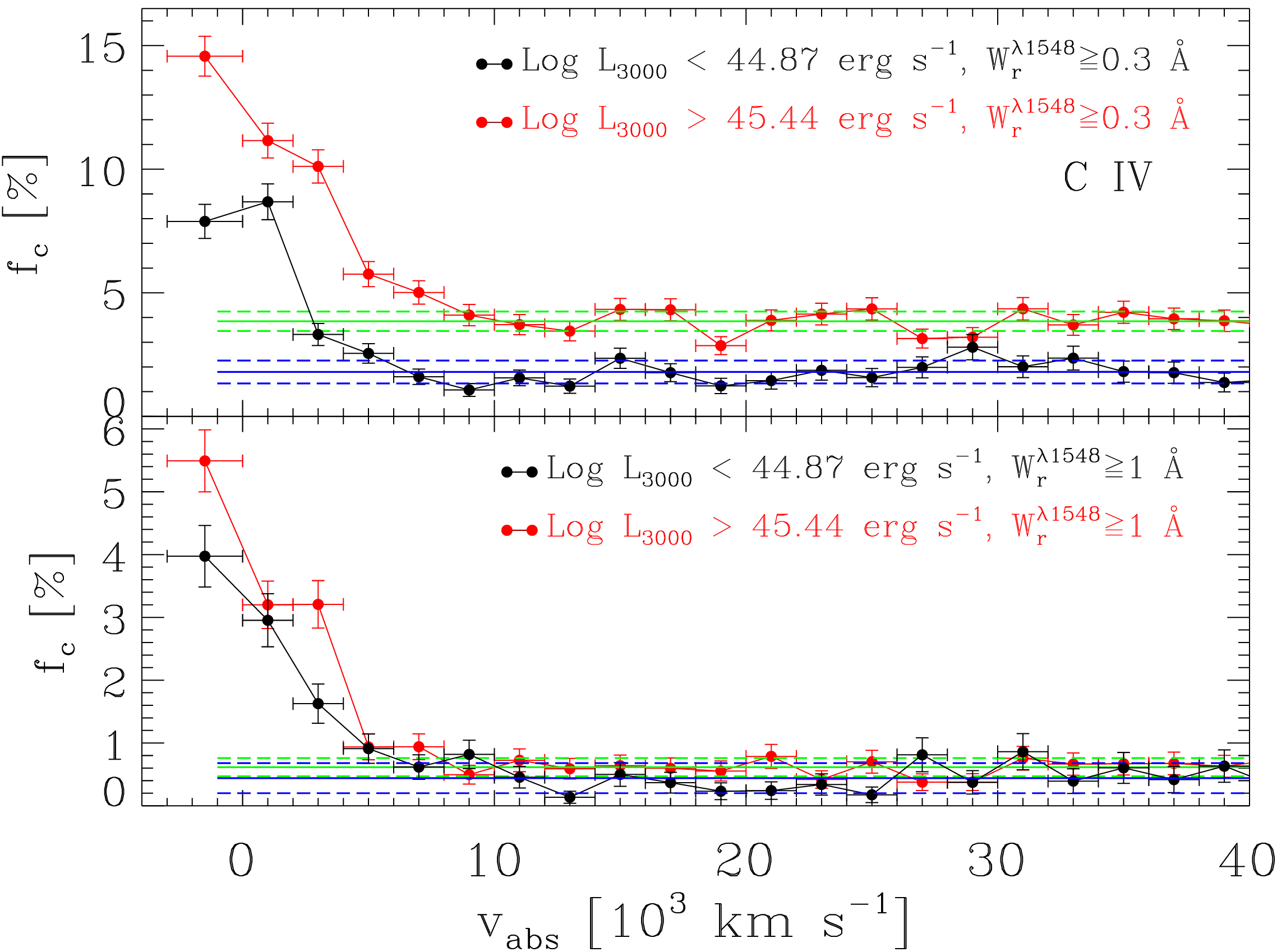}
\caption{Coverage fractions of absorptions ($f_c$) as a function of velocity offsets from the quasars. The starting \vabs\ is -3000 \kms, and the bin size is 2000 \kms\ for the absorptions with \vabs\ $>0$. Black symbols represent the low-luminosity quasars, and red symbols indicate the high-luminosity quasars. The horizonal solid-lines illustrate average coverage fractions for absorbers far away from the quasars, and the horizonal dash-lines are corresponding $\pm1\sigma$ from Poisson statistics. Left figure is for \MgII\ absorptions, and right one is for \CIV\ absorptions.}
\label{fig:cf_high_faint}
\end{figure*}

Quasar radiation could be an important mechanism driving outflow, and plays a significant role in the influence of environments. Thus, it is possible that the coverage fraction of absorbers is connected to quasar radiation. In this section, we investigate whether or not the coverage fraction of absorptions depends on quasar continuum luminosity. Here, the low- and high-luminosity quasar samples are the same as those used in Section \ref{sect:Velocity_offset}. The results are exhibited in Figure \ref{fig:cf_high_faint}. We also estimate the significance level of the excessive $f_c$ with respect to the random occurrences of intervening absorptions, which are offered in Tables \ref{tab:fcmgL} and \ref{tab:fccivL}.

\begin{table}
\caption{The significance level of coverage fraction of \MgII\ absorptions as a function of quasar luminosity.} \tabcolsep 1.mm
\centering
\label{tab:fcmgL}
 \begin{tabular}{lcccc}
 \hline\hline\noalign{\smallskip}
 \multirow{3}{*}{Velocity \kms}& \multicolumn{4}{c}{Significance level}\\
\cline{2-5}\noalign{\smallskip}
&  \multicolumn{2}{c}{$L_{\rm3000}<10^{44.05}$ \ergs} & \multicolumn{2}{c}{$L_{\rm3000}>10^{44.58}$ \ergs} \\
\cline{2-5}\noalign{\smallskip}
& \tiny{$W_r^{\lambda2796}\ge0.3$ \AA} & \tiny{$W_r^{\lambda2796}\ge1$ \AA} &  \tiny{$W_r^{\lambda2796}\ge0.3$ \AA} & \tiny{$W_r^{\lambda2796}\ge1$ \AA} \\
\hline\noalign{\smallskip}
-3000$<$\vabs$<$0   & 12.5 & 11.4 & 21.6  & 8.6\\
0$<$\vabs$<$2000    & 20.4 & 20.1 & 43.6  & 29.2\\
2000$<$\vabs$<$4000 & 1.1  & 0.7  & 7.5   & 4.7\\
4000$<$\vabs$<$6000 & 0.5  & -0.0 & 1.3   & 1.5\\
6000$<$\vabs$<$8000 & 1.1  & 1.4  & 4.1   & 4.1\\
8000$<$\vabs$<$10000& 0.5  & 0.7  & -0.9  & 2.8\\
\noalign{\smallskip}
\hline\hline\noalign{\smallskip}
\end{tabular}
\begin{flushleft}
Note --- Significance levels of excess relative to the random occurrence of intervening absorptions.
\end{flushleft}
\end{table}

\begin{table}
\caption{The significance level of coverage fraction of \CIV\ absorptions as a function of quasar luminosity.} \tabcolsep 1.mm
\centering
\label{tab:fccivL}
 \begin{tabular}{lccccc}
 \hline\hline\noalign{\smallskip}
 \multirow{3}{*}{Velocity \kms}& \multicolumn{2}{c}{Significance level}\\
\cline{2-5}\noalign{\smallskip}
&  \multicolumn{2}{c}{$L_{\rm3000}<10^{44.87}$ \ergs} & \multicolumn{2}{c}{$L_{\rm3000}>10^{45.44}$ \ergs} \\
\cline{2-5}\noalign{\smallskip}
& \tiny{$W_r^{\lambda1548}\ge0.3$ \AA} & \tiny{$W_r^{\lambda1548}\ge1$ \AA} &  \tiny{$W_r^{\lambda1548}\ge0.3$ \AA} & \tiny{$W_r^{\lambda1548}\ge1$ \AA} \\
\hline\noalign{\smallskip}
-3000$<$\vabs$<$0   & 13.2 & 14.8 & 27.2  & 33.5\\
0$<$\vabs$<$2000    & 15.0 & 10.5 & 18.5  & 17.8\\
2000$<$\vabs$<$4000 & 3.3  & 5.0  & 15.9  & 17.8\\
4000$<$\vabs$<$6000 & 1.6  & 2.0  & 4.8   & 2.2\\
6000$<$\vabs$<$8000 & -0.4 & 0.7  & 3.0   & 2.2\\
8000$<$\vabs$<$10000& -1.5 & 1.5  & 0.6   & -0.8\\
10000$<$\vabs$<$12000& -0.5 &0.1  & -0.4  & 0.8\\
\noalign{\smallskip}
\hline\hline\noalign{\smallskip}
\end{tabular}
\begin{flushleft}
Note --- Significance levels of excess relative to the random occurrence of intervening absorptions.
\end{flushleft}
\end{table}

Figure \ref{fig:cf_high_faint} and Table \ref{tab:fcmgL} clearly illustrate that the extensions of the excessive $f_c$ of \MgII\ absorptions are evidently related to quasar luminosity. For high-luminosity quasars, the excessive $f_c$ of \MgII\ absorptions with $W_r^{\lambda2796}\ge0.3$ \AA\ could be extended up to 8000 \kms, and beyond 10000 \kms\ for those with $W_r^{\lambda2796}\ge1$ \AA. While the excessive $f_c$ is mainly limited within 2000 \kms\  for low-luminosity quasars. We note in Section \ref{sect:Velocity_offset} that both associated and intervening \MgII\ absorptions would have similar detection probability. In addition, the detection probability of strong absorptions is expected to be less affected by quasar luminosity or signal-to-noise ratio of quasar spectra. Therefore, the very different luminosity dependences of the $f_c$ excess extensions can not be ascribed to signal-to-noise ratio of quasar spectra. The significantly different behaviors of $f_c$ between low-luminosity and high-luminosity quasars can be interpreted by more luminous quasars driving outflows to a higher velocity.

Turning to \CIV\ absorptions, we find that there is no evident correlation between the extension of the excessive $f_c$ and quasar luminosity. The analysis shown in Section \ref{sect:Velocity_offset} indicates that the associated \CIV\ absorptions are expected to have a slightly higher detection probability relative to the intervening ones. This could explain the slight difference of the $f_c$ excess extensions between low- and high- luminosity quasars.

\subsection{Redshift number density of absorbers}
\label{sect:dndz}
The environments of quasar-associated absorptions are expected to be different from those of intervening absorptions. Therefore, the redshift number density evolution of associated absorptions is possibly different from that of intervening absorptions. In order to further assess the excessive coverage fraction and number density ($dn/d\beta$) of absorptions, we investigate the redshift number density evolution of absorbers (dn/dz) for several subsample of absorptions.

The redshift number density of absorbers (dn/dz) is defined as the ratio between the number of absorbers and total redshift path, where the absorbers have equivalent width $W_r^{\lambda2796}\ge 0.3$ \AA\ (or $W_r^{\lambda1548}\ge 0.3$ \AA) and are located within a given velocity offset range from the quasar redshifts. The error of dn/dz can be estimated from Poisson statistics. The total redshift path covered by quasars is calculated by
\begin{equation}
\label{zpath}
 Z(W_r^{\lambda}) = \int_{z_{min}}^{z_{max}} \sum_i^{N_{spec}} g_i(W_r^{\lambda},z)dz,
\end{equation}
where $z_{min}$ and $z_{max}$ are determined by the given velocity offset range or the limits of spectra data, $g_i(W_r^{\lambda},z)=1$ if the detection threshold $W_r^{lim}\le W_r^{\lambda}$, otherwise $g_i(W_r^{\lambda},z)=0$, and the sum is over all quasars.

\begin{figure*}
\centering
\includegraphics[width=0.47\textwidth]{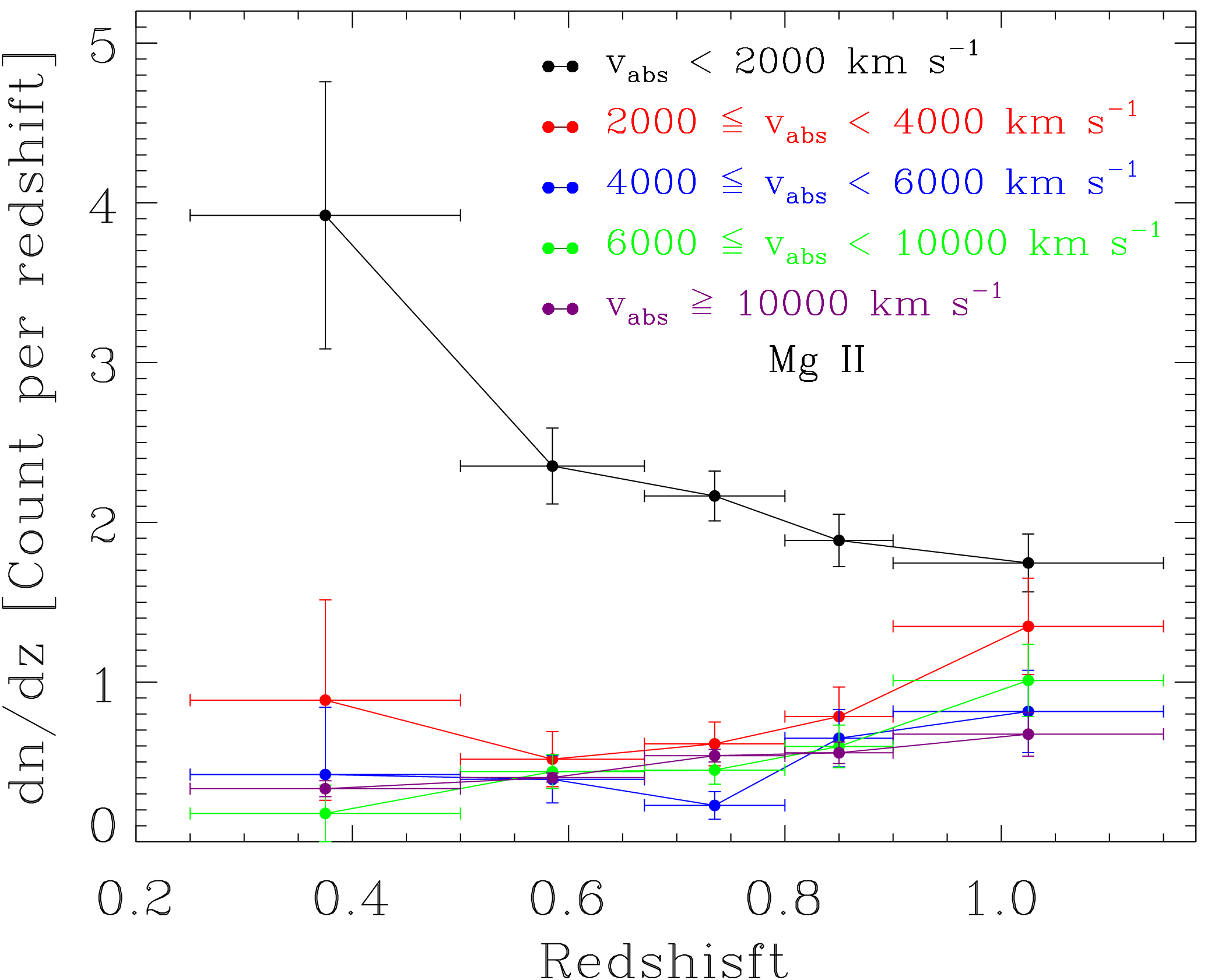}
\hspace{3ex}
\includegraphics[width=0.47\textwidth]{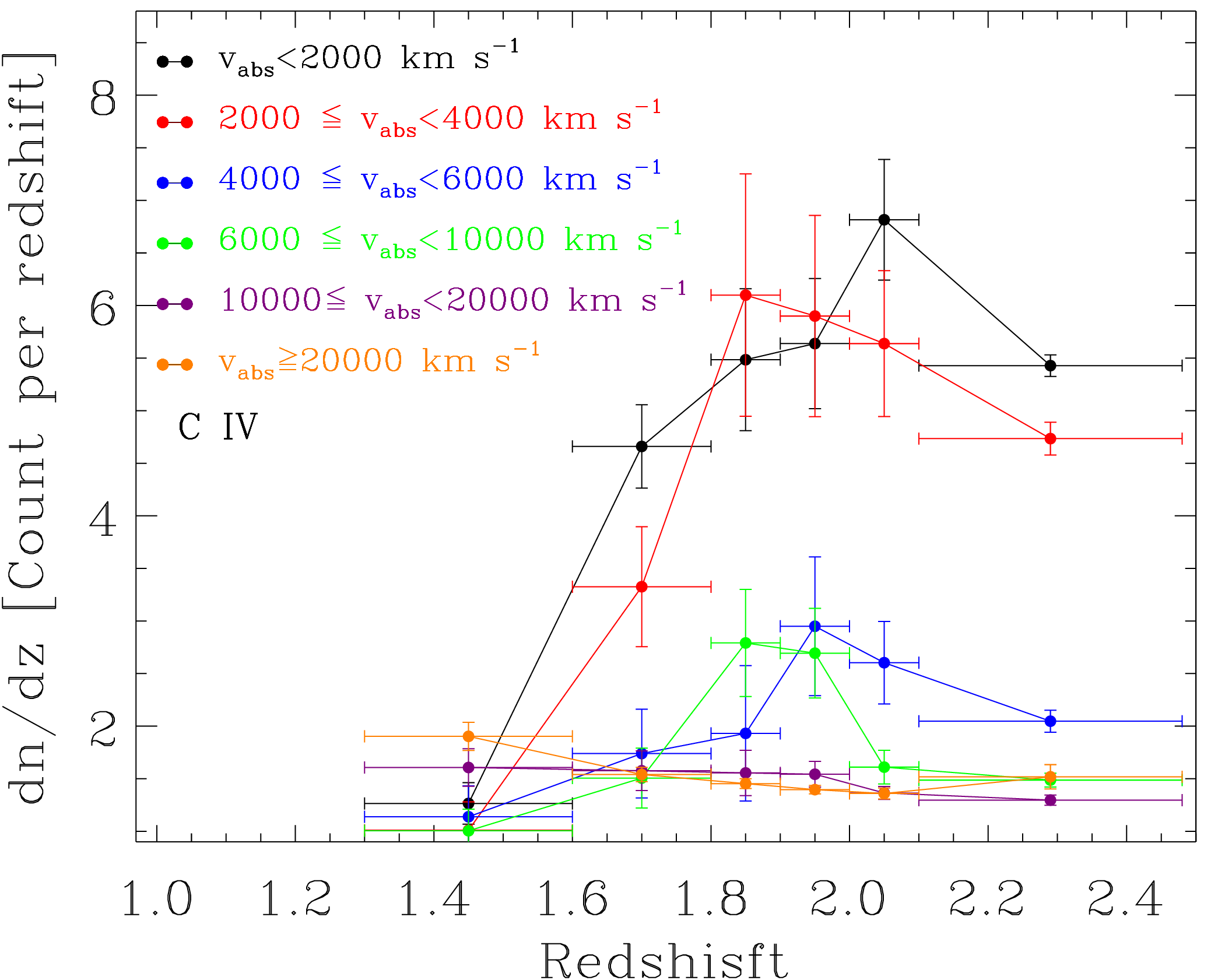}
\caption{Redshift number density evolution of absorbers. Left panel is for \MgII\ absorptions, and right one is for \CIV\ absorptions. The error bars are from the Poisson statistics. Different color symbols indicate the absorbers fallen into different velocity offset ranges from the quasars.}
\label{fig:dndz}
\end{figure*}

We investigate the evolution of dn/dz for \MgII\ absorbers in velocity offset ranges of: (1) $v_{\rm abs} <2000$ \kms, (2) $2000 \le v_{\rm abs} <4000$ \kms, (3) $4000 \le v_{\rm abs} <6000$ \kms, (4) $6000 \le v_{\rm abs} <10000$ \kms, and (5) $v_{\rm abs} \ge 10000$ \kms; and for \CIV\ absorbers in velocity offset ranges of: (1) $v_{\rm abs} <2000$ \kms, (2) $2000 \le v_{\rm abs} <4000$ \kms, (3) $4000 \le v_{\rm abs} <6000$ \kms, (4) $6000 \le v_{\rm abs} <10000$ \kms, (5) $10000 \le v_{\rm abs} <20000$ \kms, and (6) $v_{\rm abs} \ge 20000$ \kms. The results are presented in Figure \ref{fig:dndz}. Figures \ref{fig:cf_w} and \ref{fig:cf_high_faint} clearly show that the excessive $f_c$ of absorbers within 2000 \kms\ is very significant, which is possibly dominated by quasar associated absorptions, while the approximately constant $f_c$ with large velocity offset from quasars is the expectation of cosmologically intervening absorptions. We note from Figure \ref{fig:dndz} that the dn/dz of \MgII\ absorbers with $v_{\rm abs} <2000$ \kms\ evidently differs from that of \MgII\ absorbers with $v_{\rm abs} \ge 4000$ \kms, and the dn/dz of \CIV\ absorbers with $v_{\rm abs} <4000$ \kms\ obviously differs from that of \CIV\ absorbers with $v_{\rm abs} \ge 10000$ \kms. The violent activity and strong radiation within quasar center, and their powerful feedback to surrounding environment likely drive an evolution behaviour of $dn/dz$ for associated absorptions which obviously differs from that of intervening absorptions. These different dn/dz evolutions between the absorbers with low and high velocity offset from the quasars would be ascribed to different environments in which the absorbers reside. That is, the absorbers with low velocity offset are physically associated with quasar systems and those with high velocity offset are connected to cosmologically intervening medium, though there are a few evidences of NAL outflows with very high velocity \cite[e.g.,][]{2004ApJ...601..715N,2013MNRAS.434..163H,2011MNRAS.410.1957H,2013MNRAS.434.3275C,2013ApJ...777...56C,2015MNRAS.450.3904C}. The $dn/dz$ evolution of \MgII\ absorbers with $2000 \le v_{\rm abs} <4000$ \kms, and that of \CIV\ absorbers with $4000 \le v_{\rm abs} <10000$ \kms\ are possible between the $dn/dz$ behaviours of associated and cosmologically intervening absorbers. These intermediate states suggest that therein some absorbers would be truly associated with quasars and some are located within cosmologically intervening environments. Relative to the very significantly excessive $f_c$ of the associated absorbers, the excessive $f_c$ of the absorbers in the intermediate states (see Figure \ref{fig:cf_w} and \ref{fig:cf_high_faint}, and Tables \ref{tab:fcmg}, \ref{tab:fcciv}, \ref{tab:fcmgL} and \ref{tab:fccivL}) is much less significant. This indicates that a considerable portion of the absorbers in the intermediate states would be not truly associated with the quasar systems.

\begin{figure}
\centering
\includegraphics[width=0.47\textwidth]{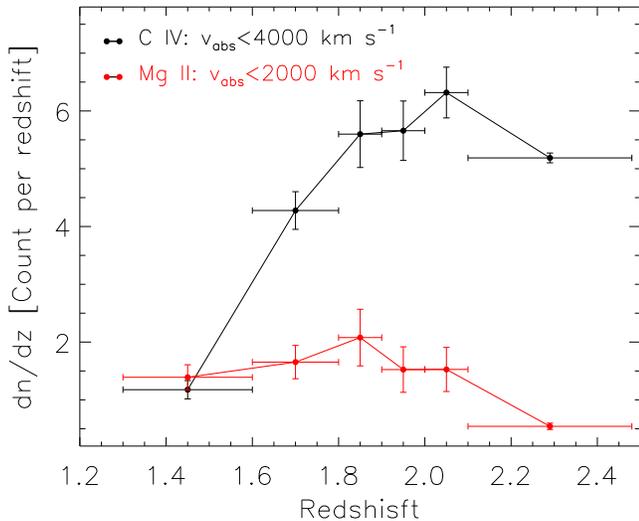}
\caption{Redshift number density evolution of associated \CIV\ (black symbols) and \MgII\ (red symbols) absorbers. Both \CIV\ and \MgII\ absorptions are detected from the same \CIV\ quasar sample.}
\label{fig:dndzciv_mg}
\end{figure}

Previous works \cite[e.g.;][]{2012ApJ...761..112M,2013ApJ...770..130Z,2013ApJ...763...37C,2015ApJS..221...32C,2016MNRAS.462.2980C} have revealed that the number density evolution of intervening absorptions for both \MgII\ and \CIV\ are similar to the cosmic star formation history \cite[e.g.;][]{2014ARA&A..52..415M}. In other words, the incidences of the intervening \CIV\ and \MgII\ absorptions have a parallel evolution behaviours. Figure \ref{fig:dndz} shows that the number density evolution of \MgII\ absorptions with $v_{\rm abs} > 4000$ \kms\ sightly differs from that of \CIV\ absorptions with $v_{\rm abs} > 10000$ \kms, which would be due to the evolution effect since they are located within different redshift ranges. While, Figure \ref{fig:dndz} also exhibits that the redshift number density evolutions of absorptions are very different between \MgII\ with \vabs$<2000$ \kms\ and \CIV\ with \vabs$<4000$ \kms. This significant difference could not be ascribed to the evolution effect. In order to reduce the evolution effect, it would better to compare the evolution behaviours of associated \MgII\ and \CIV\ absorptions within the same redshift range. Here, we directly adopt the \MgII\ absorption systems with $v_{\rm abs} < 2000$ \kms\ included in the absorption catalog of \cite{2015ApJS..221...32C} for the quasars contained in \CIV\ quasar sample. Figure \ref{fig:dndzciv_mg} exhibits the incidences of the absorptions for \MgII\ with $v_{\rm abs} < 2000$ \kms\ and for \CIV\ with $v_{\rm abs} < 4000$ \kms. It is clear that the redshift number density evolution of the \MgII\ associated absorptions significantly differs from that of the \CIV\ associated absorptions, which might be related to ionization potentials of ionized gas and strong quasar radiation field. The fundamental origins of this different redshift dependences will be detailedly discussed in our future work.

\section{Conclusions and summary}
\label{sect:summary}
Using large samples of the SDSS quasar \CIV\ and \MgII\ narrow-line absorptions \cite[][]{2015ApJS..221...32C,2016MNRAS.462.2980C} that include both associated and intervening components, we mainly investigate absorber properties in three ways: (1) the distribution of the number density of absorbers per velocity offset from quasar redshifts ($dn/d\beta$); (2) the distribution of coverage fraction ($f_c$) that is the ratio between the number of quasars, which exhibit at least one absorber with an equivalent width at rest-frame $\ge0.3$ \AA\ and significance level $\ge3\sigma$, and the total number of quasars that can be used to detect absorptions with given criteria; and (3) the redshift number density of absorbers ($dn/dz$). We find that all the $dn/d\beta$, $f_c$, and $dn/dz$ show very different distributions for the absorbers with low velocity offsets from the quasars and the intervening absorbers. The main results are as follows.

\begin{description}
  \item[(1)] Both $dn/d\beta$ and $f_c$ distributions of \MgII\ absorptions are related to absorption strength and quasar continuum luminosity. Both the significantly excessive $dn/d\beta$ and $f_c$ relative to the random occurrence of intervening absorptions are limited within 2000 \kms\ when quasar continuum luminosity is $L_{\rm 3000} < 10^{44.05}$ \ergs\ or absorption strength is $W_r^{\lambda2796}\ge2$ \AA. While, the evident excesses can be extended to 4000 \kms\ for the middle strength absorptions with $W_r^{\lambda2796}\ge1$ \AA, and beyond 8000 \kms\ for the weak absorptions with $W_r^{\lambda2796}\ge0.3$ \AA\ and for the luminous quasars with $L_{\rm 3000} > 10^{44.58}$ \ergs. The excessive tails for the luminous quasars and for the weak absorptions are much farther than 3000 \kms, which is the traditional cut to divide associated and intervening \MgII\ absorptions. However, the significant level of the excesses beyond 2000 \kms\ is much less than that within 2000 \kms. In addition, we find that the redshift number density evolution of \MgII\ absorbers with \vabs\ $<2000$ \kms\ evidently differs from that with \vabs\ $>2000$ \kms\ though there is a slight cross in velocity offset range of $2000 \sim 4000$ \kms. The weak excess of both the $dn/d\beta$ and $f_c$ beyond 2000 \kms, and the slight cross of the $dn/dz$ evolution in $2000<v_{\rm abs}<4000$ \kms\ could be mainly originated from a few \MgII\ outflows with high velocity, which would not evidently contribute to the fraction of quasar associated absorptions. Therefore, we suggest a velocity offset cut of 2000 \kms\ to define quasar associated systems of \MgII\ narrow absorption lines.
  \item[(2)] Both the relative excess $dn/d\beta$ and $f_c$ of \CIV\ absorptions is neither related to absorption strength nor to quasar continuum luminosity. The relative excesses can be extended up to 10000 \kms. We note that the relative excesses are mainly concentrated within \vabs$<4000$ \kms, which are much more significant than that beyond 4000 \kms. In addition, the absorbers with \vabs$<4000$ \kms\ show obviously different redshift number density evolution from those with \vabs$>10000$ \kms. The $dn/dz$ evolution of absorbers with $4000<v_{\rm abs}<10000$ \kms\ is between those with \vabs$<4000$ \kms\ and \vabs$>10000$ \kms. For the absorbers with $4000<v_{\rm abs}<10000$ \kms, the weak relative excesses of both the $dn/d\beta$ and $f_c$, and the intermediate state of the redshift number density evolution would be mainly come from a few \CIV\ outflows with high velocity. It is comparative safe to define quasar associated systems of \CIV\ narrow absorption lines with a velocity cut of 10000 \kms\ considering a few outflows with high velocity. While, with respect to the very significant excess of both the $dn/d\beta$ and $f_c$ within 4000 \kms, the much weaker excess beyond 4000 \kms\ indicates that the high velocity outflows would not evidently reduce the fraction of quasar associated systems of \CIV\ narrow absorption lines. Combining with the obvious difference of redshift number density evolution between absorbers with \vabs$<4000$ \kms\ and \vabs$>10000$ \kms, we suggest a velocity offset cut of 4000 \kms\ to define quasar associated systems of \CIV\ narrow absorption lines.
\end{description}

\acknowledgements We are grateful to the anonymous referee for comments that help to improve this manuscript. This work was supported by the National Natural Science Foundation of China (NO. 11363001; NO. 11763001), and the Guangxi Natural Science Foundation (2015GXNSFBA139004).

Funding for SDSS-III has been provided by the Alfred P. Sloan Foundation, the Participating Institutions, the National Science Foundation, and the U.S. Department of Energy Office of Science. The SDSS-III web site is http://www.sdss3.org/.


\end{document}